\documentclass[12pt]{article}
\pdfoutput=1
\usepackage{Max}
\def\b{\beta}
\def\chb{\Hh(\b')}
\def\chob{\Hh_1(\b')}
\def\pchb{\mathbb{P}\chb}
\def\ph{\mathbb{P}\Hh}
\newcommand\ignore[1]{}

\renewcommand{\l}{\lambda}
\newcommand\eq[1]{\begin{align}#1\end{align}}

\def\frakC{\mathfrak{C}}
\usepackage{mathtools}
\def\fracC{\mathfrak{C}}

\def\calW{\mathcal{W}}
\usepackage{etoolbox}
\AtBeginEnvironment{quote}{}

\def\core{\operatorname{Core}}
\def\dcore{\partial\text{Core}}
\def\icore{\mathcal{I}\core}
\def\cs{\core(\Sigma)}
\def\dcs{\dcore(\Sigma)}
\def\ics{\icore(\Sigma)}
\def\ths{\Theta(\Sigma)}
\def\ks{k(\Sigma)}
\usepackage{subcaption}

\title{Semiclassical Entropy of BPS States in 4d $\mathcal{N}=2$ Theories and Counts of Geodesics}
\author[1]{Shamit Kachru\thanks{skachru@stanford.edu}}
\author[1]{Richard Nally\thanks{rnally@stanford.edu}}
\author[2]{Arnav Tripathy\thanks{tripathy@math.harvard.edu}}
\author[1]{Max Zimet\thanks{mzimet@fas.harvard.edu}}

\affil[1]{Stanford Institute for Theoretical Physics,

Stanford University, Stanford, CA 94305 USA

~}
\affil[2]{Department of Mathematics,

Harvard University, Cambridge, MA 02138 USA}

\date{}

\begin{document}

\maketitle

\begin{abstract}
We relate a number of results in the theory of flat surfaces to BPS spectra of a class of 4d $\N=2$ supersymmetric quantum field theories arising from M5 branes wrapped on Riemann surfaces -- $A_1$ class S theories.  In particular, we apply classic results of Eskin and Masur, which determine the asymptotic growth of geodesic counts at large length on flat surfaces, as well as more recent progress in the mathematics literature, to determine the large mass asymptotics of the BPS spectra of a wide class of such theories at generic points in the Coulomb branch.

\end{abstract}

\newpage
\tableofcontents
\hypersetup{linkcolor=blue}

\section{Introduction}

The BPS states of a theory with extended supersymmetry play a crucial role in understanding its physics. The solution of 4d $\N=2$ supersymmetric field theories 
\cite{sw:theory1, sw} provides
a clear demonstration of this.  In addition to elucidating the strong-coupling dynamics of a class of non-Abelian gauge theories, such solutions also allowed
the discovery of novel fixed points of the renormalization group, starting with the work \cite{argyresDouglas}.  String theory fits into the story in a very natural way.  In one 
avatar,
string dualities (including heterotic/type II duality and mirror symmetry) allow for a solution of quantum $\N=2$ gauge theories to emerge from {\it classical} string theory \cite{kachru:heteroticK3T2,kachru:geomEng}.  This technology can be developed into a technique of ``geometric engineering" which allows, in principle, for the determination
of the BPS spectrum and other properties of a wide class of $\N=2$ theories \cite{vafa:geodesics, vafa:geoEng, vafa:sol}.  
In another avatar, realizing the gauge theories on M5-branes gives rise to a wide class of theories and makes manifest some striking dualities
\cite{w:MGaugeSol,gaiotto:classS, GMN:classS}.  Further work on $\N=2$ theories has proceeded in many directions following these developments.

The goal of this paper will be to connect a recent line of research in mathematics to the problem of determining the spectrum of BPS states in 4d $\N=2$ field theories.
We will particularly be concerned here with determining the {\it asymptotic growth} of such counts, which can be relevant for problems ranging from studies of holography and black hole entropy to the determination of metrics on hyper-K\"ahler manifolds to the study of thermodynamic phenomena in quantum field theories (with \cite{mz:k3,moore:wildWall} providing some recent applications).  
We proceed via the observation, made long ago in the physics literature, that such states can be related to geodesics on the Seiberg-Witten curve \cite{vafa:geodesics,brand:geodesics,warner:geodesics,rabin:geo,vafa:AD,GMN:classS}.  Building on the recent discussion in \cite{mz:slagCounts}, we connect this
problem to studies of dynamics on flat surfaces in the mathematics literature, beginning with classic work of Masur, Eskin and others -- for nice reviews see
\cite{zorich:flat,eskin:notes}.  Our main point is that both this old work and more recent work of mathematicians (appearing in papers such as \cite{tahar:poles, tahar:twoStar, aulicino:cbrank}), as well as conjectured extensions, allows for the determination
of BPS asymptotics of a wide class of $\N=2$ theories. These results, in turn, make interesting predictions for the variety of other problems that have been related to BPS state counting (see, e.g., \cite{yi:surfaces,mikhailov:surface,sethi:FWebs,bergman:FWebs,zwiebach:webs,GMN:classS,bs:quadStab,GMN:networks,fiol:fractBrane,denef:hall,gukov:refineMotive2,vafa:RTwist,vafa:complete,vafa:quiver1,vafa:quivers,Chuang:2013wt,
longhi:invt,longhi:quiverGraph,moore:semiclassical}).

The organization of this paper is as follows.  In \S\ref{sec:max} we give a reasonably detailed review of class S $\N=2$ field theories, with a focus on the $A_1$ theories our techniques
allow us to study. Next, in \S\ref{sec:geo}, we describe the results coming from mathematicians' work on geodesics of flat surfaces. In \S\ref{sec:geoA1}, we show that, frabjously, these results allow us to
determine the asymptotic behavior of BPS spectra of an infinite number of $A_1$ class S theories. Finally, we conclude in \S\ref{sec:conclude}.

\section{Theories of Class S and their BPS Spectra}
\label{sec:max}
\subsection{Stringy constructions}

We begin by reviewing a number of aspects of 4d $\N=2$ field theories of class S. Much of this material is contained in the review \cite{tachikawa:N2}, to which we refer readers who seek more details on 4d $\N=2$ field theories.

Class S theories are 4d $\N=2$ field theories obtained via twisted compactification of $(2,0)$ theories on a (possibly punctured) Riemann surface $C$, which we will call the ultraviolet curve. We denote by $\bar C$ the compactification of $C$. Throughout this paper, we restrict to the $A_1$ $(2,0)$ theory associated to the $\mathfrak{su}(2)$ Lie algebra, so the data specifying our theories is a choice of $C$ and a further choice of some data associated to the codimension two defects located at the punctures.

Embedding this construction into string theory will prove useful. We do so in two ways. One is by compactifying a pair of M5-branes on the zero section $C\subset T^*C$ and taking the low energy limit to obtain a 4d field theory. We will be interested in the Coulomb branch, $\B$ (which we parametrize by a coordinate $u$), of this field theory, which describes the M5-branes puffing up into a single M5-brane wrapping a holomorphic curve $\Sigma\subset T^*C$ -- the Seiberg-Witten curve\footnote{This mathematical characterization of the Seiberg-Witten curve traces back in the physics literature at least to \cite{gorsky:integrable,martinec:integrable,w:integrable}.} -- which is a branched double cover of the zero section $C$. The twist is implemented by the non-trivial normal bundle of $\Sigma$ \cite{vafa:k3inst}.

Throughout this paper, we let $z$ be a holomorphic local coordinate on $C$ and $\lambda = x \, dz$ an element of the fiber $T^*_zC$ over $z$. $\lambda$ pulls back trivially to $T^*C$ and defines a canonical one-form thereon (that is, an element of $T^*(T^*C)$), which via a mild abuse of notation we will also call $\lambda$. We then obtain a canonical holomorphic symplectic form $\omega = d\lambda = dx\wedge dz$ on $T^*C$. This is equivalent to $T^*C$ being hyper-K\"ahler.\footnote{Actually, this is an oversimplification -- as stated, this implication is only known to be correct for compact manifolds. This is related to the fact that the Calabi-Yau theorem only applies to compact manifolds. $T^*C$ has only been demonstrated to be hyper-K\"ahler in some cases, and examples of compact complex manifolds $C$ for which $T^*C$ has no complete hyper-K\"ahler metric are known \cite{feix:cotangent}. On the other hand, it is also known that for any K\"ahler manifold $C$ an open neighborhood of the zero section $C\subset T^*C$ is hyper-K\"ahler \cite{kaledin:hk,kaledin:improve,feix:cotangent}, and this may suffice for many physical applications. Anyways, if we are to be more careful, we should perhaps think of $T^*C$ as arising in a gravitational decoupling limit of a hyper-K\"ahler manifold \cite{GMN:classS}.} So, compactification of M-theory on $T^*C$ preserves 16 supercharges, and the worldvolume of an M5-brane wrapping a holomorphic curve $\Sigma\subset T^*C$ preserves 8 supercharges. By restricting $\lambda$ to $\Sigma$, we obtain a meromorphic differential on the compactification $\bar\Sigma$ of $\Sigma$ (which by a further abuse of notation we will also call $\lambda$), called the Seiberg-Witten differential.

Punctures in this picture are associated to additional M5-branes. Intuitively, they fill the fibers of $T^*\bar C\to \bar C$ above the punctures of $C$, as well as the four dimensions associated to the 4d field theory. They preserve the 8 supercharges of our compactification, and so we still have a 4d $\N=2$ theory. More precisely, after merging with our original M5-branes they enforce certain boundary conditions on $\Sigma$.

The Seiberg-Witten curve is defined by an equation\footnote{\eqref{eq:swPhi} makes clear that $\Sigma\to C$ is a `canonical double cover,' in the sense of \cite{strebel:quad,chen:kDiff}.}
\be \lambda^2 = \phi \ , \label{eq:swPhi} \ee
or equivalently
\be x^2 = \varphi \ , \label{eq:swU} \ee
where $\phi = \varphi(z) \, dz^2$ is a quadratic differential on $C$ (that is, a section of $(T^*C)^{\otimes 2}$) which depends on the parameters and moduli of the field theory. To stress the dependence of $\Sigma$ on $u$, we will sometimes denote the curve as $\Sigma_u$. The data associated to the punctures constrain the singular terms that $\phi$ may have there. Away from these points, $\phi$ must be holomorphic. The Coulomb branch (including singular points) is always topologically an affine subspace of a vector space of the form
\be H^0\parens{C,(T^*C)^{\otimes 2}\parens{\sum p_i z_i}} \ ; \label{eq:affine} \ee
$(T^*C)^{\otimes 2}\parens{\sum_i p_i z_i}$ denotes the bundle of quadratic differentials twisted by the line bundle dual to the divisor $\sum_i p_i z_i$, whose sections are quadratic differentials which are holomorphic everywhere except for possibly at the punctures $z_i$ where they may have a pole of order at most $p_i$. We classify poles as simple (order 1), regular (order at most 2), and irregular (order greater than 2). The choice of the affine subspace involves the mass parameters and dynamically generated scale(s), which constrain the coefficients of $\phi$. (One can of course set one dynamical scale to unity by a choice of units, but the associated constraint on $\phi$ remains. Such constraints can be present even in superconformal field theories; indeed, we take $z$ to be dimensionless, while $\lambda$ has dimensions of mass, so in such theories certain terms in $\phi$ may require a factor of $\Lambda^2$, where $\Lambda$ is an arbitrary scale with no effect on physics.) We denote the complex dimension of $\B$ by $r$; in Lagrangian theories, this will be the rank of the gauge group.

Note that the zeroes of $\phi$ are generically all simple; indeed, when this is not the case, a BPS state becomes massless (as will be clear from the geometric picture of BPS states introduced below). A similar phenomenon occurs when a zero collides with a simple pole. By considering these phenomena in tandem, we define the singular locus $\B^{\rm sing}\subset \B$, which we henceforth avoid, to be the subset of the Coulomb branch where smooth branch points of $\bar\Sigma\to \bar C$ collide. (That simple poles yield branch points, which correspond to smooth points of $\bar\Sigma$, will be demonstrated in \S\ref{sec:rels}. We exclude these points from $\Sigma$, so that every point of $\Sigma$ has an image in $C$.)

A second method for constructing 4d $\N=2$ field theories in string theory is geometric engineering -- that is, compactification of type II string theories on non-compact Calabi-Yau threefolds in a decoupling limit where the threefold becomes singular as $\alpha' \to 0$. Indeed, this approach can produce some theories that do not appear to be of class S \cite{vafa:sol}, but we will focus on $A_1$ class S theories. In type IIB, we can realize these by compactifying in two steps. First, we compactify on an $A_1$ singularity ($\CC^2/Z_2$) in order to obtain a T-dual description of the $A_1$ $(2,0)$ theory to the one involving two parallel M5-branes. Deformation of this singularity corresponds to the chiral operator which will give rise to the Coulomb branch upon dimensional reduction -- that is, the topological twist turns this operator into a quadratic differential (whose expectation value we have denoted by $\phi$). (After all, in type IIB vector multiplet scalars correspond to complex structure deformations.) Then, we implement the desired twisted compactification on $C$ by fibering deformed $A_1$ singularities over $C$. Punctures are simply singularities in $C$. To associate equations to this picture, recall first that the defining equation of the deformed $A_1$ singularity is\footnote{To see this, we make the change of variables $\alpha = y + i w, \beta = -y+iw$ and undeform the singularity by setting $\varphi=0$. This yields $\alpha\beta = x^2$. Then, writing $\alpha=r^2,\beta=s^2,x=rs$ gives the vacuous condition $r^2s^2=r^2s^2$, which identifies the variety $\alpha\beta=x^2$ with $\CC^2/Z_2=\{(r,s)\in \CC^2\}/\avg{-1}$.\label{ft:singRes}}
\be w^2 + x^2 + y^2 = \varphi \ , \label{eq:deformedA1} \ee
where $\varphi$ is a constant. Our threefold geometries then take the form \eqref{eq:deformedA1}, where $\varphi$ is promoted to a function of $z$; more precisely, we carve out a threefold from the total space of $(T^*C)^{\oplus 3}$, parametrized by $z$ and one-forms $w'=w\, dz,y'=y\, dz,\lambda=x\, dz$, via the equation
\be w'^2 + y'^2 = \phi - \lambda^2 \ , \label{eq:promoted} \ee
where $\phi=\varphi(z)\, dz^2$ is a quadratic differential on $C$.\footnote{More generally, motivated by footnote \ref{ft:singRes}, we can take $r$ and $s$ to be sections of arbitrary line bundles $L_1$ and $L_2$ such that $L_1\otimes L_2=T^*C$. Then,
\be \alpha\otimes\beta = \lambda^2 - \phi \ee
carves out a threefold from the total space of $L_1^{\otimes 2}\oplus L_2^{\otimes 2}\oplus (L_1\otimes L_2)$, where the three summands are respectively associated to $\alpha,\beta,\lambda$. This yields the approach in the text by choosing $L_1$ and $L_2$ to be the same square root of $T^*C$ and defining $\alpha = y'+iw', \beta=-y'+iw'$. These different choices of $L_1$ and $L_2$ seem unlikely to affect the field theory obtained in the decoupling limit -- at the very least, the IR physics and BPS spectra seem indifferent to these choices.} The holomorphic 3-form is
\be \Omega = - \frac{1}{2\pi} \frac{dx \wedge dy \wedge dz}{w} \ .\footnote{This may be obtained from the 4-form $dw\wedge dx\wedge dy\wedge dz$ via the Poincar\'e residue.} \label{eq:3form} \ee
To discover the role of $\Sigma$ in this geometry, we first note that the deformed $A_1$ singularity has a non-trivial 2-sphere. This follows by defining $\tilde w=w/\sqrt{\varphi},\tilde x=x/\sqrt{\varphi},\tilde y=y/\sqrt{\varphi}$, so that
\be \tilde w^2 + \tilde x^2 + \tilde y^2 = 1 \ .\ee
The 2-sphere is then simply obtained by taking $\tilde w,\tilde x,\tilde y$ to be real. We think of $\tilde x$ as defining latitude on the sphere, with $\tilde x=\pm 1$ defining the north and south poles. But, this is simply the equation \eqref{eq:swU} for $\Sigma$! So, the two sheets of $\Sigma$ correspond to the poles $y=w=0$ of this sphere, fibered over $C$. For later reference, we note that the integral of $\Omega$ over a latitude is given by
\be \int \Omega = \omega \ , \label{eq:latInt} \ee
where $\omega = dx\wedge dz$ is the 2-form defined in the M-theory construction.

\subsection{Geometry of BPS states}

So far, we described two methods of constructing $\N=2$ theories in string theory. 
 There are two aspects of these constructions that we wish to highlight: geometric realizations of BPS states and the relationship between this picture and the low-energy description of physics on the Coulomb branch. First, we turn to the geometric description of BPS states. In the M-theory picture, these are M2-branes wrapping special Lagrangian 2-cycles in $T^*C$ that terminate on $\Sigma$ \cite{yi:surfaces,mikhailov:surface,sethi:FWebs}, while in the type IIB picture they are D3-branes wrapping special Lagrangian 3-cycles. Since the latter are calibrated by $\Omega$ (times a phase), we can determine their volume by integrating $\Omega$ over any homologous 3-cycle. Such a 3-cycle will roughly be a combination of the 2-sphere discussed above and a 1-cycle on $C$. To compute the volume of this 3-cycle, we first integrate over the latitude; \eqref{eq:latInt} yields the 2-form $\omega$ which is to be integrated over an effective \emph{2-brane} worldvolume. This 2-brane appears to be stretched in one dimension between the two sheets of $\Sigma$ and fibered over $C$ in the other dimension. This is precisely the mathematical description of the volume of a special Lagrangian M2-brane ending on $\Sigma$ in the M-theory frame \cite{vafa:geodesics}! This relationship between D3-branes and M2-branes follows simply from the T-dual constructions of M-strings in the $A_1$ $(2,0)$ theory as either M2-branes stretched between M5-branes or D3-branes wrapping the 2-sphere in the $A_1$ ALE space. For, the BPS states in the 4d class S theories are obtained by wrapping these strings on 1-cycles as we compactify.

Indeed, we can benefit from focusing further on these strings \cite{vafa:geodesics}. To do so, we compute the tension of an M-string constructed by stretching an M2-brane between the intersections of the two sheets of $\Sigma$ with some fiber $T^*_zC$:
\be \int \omega = \lambda_+ - \lambda_- \ .\ee
Here, $\lambda_\pm$ is the Seiberg-Witten differential restricted to the two sheets of $\Sigma$. We thus find that if the boundary of the M2-brane in $\Sigma$ is the (closed) union of 1-cycles $\gamma=\gamma_+\cup\gamma_-$ in the two sheets of $\Sigma$, then the mass of the associated BPS state is
\be M_\gamma = \abs{\int_{\gamma_+}\lambda + \int_{\gamma_-} \lambda} = \abs{\int_\gamma\lambda} \ . \ee
Said another way, the central charge of a BPS state associated to the 1-cycle $\gamma$ is
\be Z_\gamma = \int_\gamma \lambda \ . \label{eq:ZSW} \ee
This M2-brane picture demonstrates that the (flavor plus gauge) charge lattice $\hat\Gamma_u$ of the IR $U(1)^r$ gauge theory is geometrically the relative homology group $H_2(T^*C,\Sigma_u;\ZZ)$. In terms of the M-string picture, this is the subset of $H_1(\Sigma_u;\ZZ)$ that is odd under the deck transformation which exchanges the two sheets of $\Sigma_u$ (or equivalently which is even under a deck transformation followed by reversal of the 1-cycle) \cite{GMN:classS}. (The cycles we created by deleting smooth points of $\bar\Sigma$ which correspond to simple poles of $\phi$ are even under the deck transformation, and so do not alter the charge lattice.) Similarly, the gauge charge lattice $\Gamma_u$ is the analogous subset of $H_1(\bar\Sigma_u;\ZZ)$, where $\bar\Sigma_u$ is the compactification of $\Sigma_u$ obtained by filling in its punctures.\footnote{These characterizations of the charge lattices are not entirely accurate. One reason is that, as we will see below, there may be flavor charges which exist, but to which there is no associated mass parameter, and these charges will be absent from $\hat\Gamma_u$ and $\Gamma_{\rm flavor}$. Another is that there are flavor charges associated to simple poles of $\phi$, but these flavor lattices only notice them when we turn on the associated mass parameters. The geometric characterization of BPS states in the text is still complete; that is, there is a more refined grading of BPS states by flavor charges than is evident from the class S construction, but at least we are not missing states.} The Dirac-Schwinger-Zwanziger symplectic pairing on $\hat\Gamma_u$ is simply the restriction of the intersection form $\avg{,}$ on $H_1(\Sigma_u;\ZZ)$ to $\hat\Gamma_u$; its radical is the flavor charge lattice $\Gamma_{\rm flavor}$, which does not depend on $u$.

We now make further use of the special Lagrangian condition, which implies that for some phase $e^{i\theta}$, $\lambda e^{i\theta}$ is the effective volume form for a BPS state in class $\gamma$. Therefore,
\be M_\gamma = \abs{\int_\gamma\lambda} = \int_{\tilde\gamma} ds \ ,\label{eq:massintegral} \ee
where we have defined the metric
\be ds^2 = \Real \lambda\otimes \bar\lambda \label{eq:metric} \ee
on $\Sigma_u$, with respect to which $\tilde\gamma$ is a geodesic in homology class $\gamma$ along which
\be \iota_{\partial_s}\lambda \in e^{-i\theta}\RR^+ \ . \label{eq:geodesicAngle} \ee
($\iota$ denotes interior product, or contraction.) Note that away from zeroes and punctures this metric is flat, since we can choose a local holomorphic coordinate so that $\lambda=dw$. We thus arrive at the study of geodesics on flat surfaces, which will occupy much of this paper. We emphasize that this is not the familiar constant curvature metric on Riemann surfaces; indeed, we have conical singularities at the zeroes of $\lambda$.

BPS geodesics contribute entire supersymmetry multiplets to the 4d field theory; the specific multiplets a priori depend on details of the geometry. Remarkably, in $A_1$ class S theories the following simple topological classification suffices: M2-brane discs (with a single boundary component) contribute a hypermultiplet, cylinders contribute a vector multiplet, and no other topologies give BPS states \cite{yi:surfaces,mikhailov:surface}. Equivalently, in the IIB language these are D3-branes with topology $S^3$ (which should be thought of as the canonical 2-sphere fibered over an interval, at whose ends it shrinks to a point) and $S^2\times S^1$, respectively \cite{vafa:geodesics}. To determine these results one must determine the supersymmetric states of M2-brane or D3-brane worldvolume theories. This is equivalent to determining the BPS states of supersymmetric quantum mechanical non-linear sigma models with targets the moduli spaces of the cycles wrapped by these branes. In the type IIB picture, the dimension of these moduli spaces is given by the first Betti number of the cycle \cite{mclean:dim}. An analogous result holds in the M-theory picture, where the first Betti number $2\tilde g+\tilde b-1$ now receives contributions from both the genus $\tilde g$ and the number $\tilde b$ of boundary components \cite{mikhailov:surface}. By this dimension counting, cycles with the hyper topology ($S^3$ or a disc, depending on the picture) are rigid; including the universal center of mass moduli gives a hyper. In contrast, cycles with the other topology have a one-dimensional moduli space; indeed, it is an interval (which in IIA brane engineering is the $x^6$ interval between NS5-branes).\footnote{To picture this interval, imagine a closed loop wrapping the handle of a genus 2 surface between the two holes. This picture also helps emphasize the fact that the metric \eqref{eq:metric} is not the familiar one with constant negative curvature, as every geodesic in this moduli space has the same length. Finally, we note an exception to the rule that the moduli space is an interval: a closed geodesic on an elliptic curve has an $S^1$ moduli space. This makes sense, since this geometry must give rise to an $\N=4$ $1/2$-BPS vector multiplet. The change in the multiplet is due both to the change in the target space of the quantum mechanics and the fact that we have $\N=8$ quantum mechanics instead of $\N=4$.} Quantizing this moduli space (and including center of mass moduli) gives a vector multiplet \cite{mikhailov:surface}. The constraints on special Lagrangian cycles are sufficiently restrictive that generically these are the only topologies that give BPS states \cite{yi:surfaces}. The geodesic picture makes this clear \cite{GMN:classS}.

This can all be translated to the geodesic picture. Hypers correspond to saddle connections -- that is, geodesics on $\bar\Sigma$ between distinct smooth branch points of $\bar\Sigma\to \bar C$ (which avoid such points except for at the endpoints); to get a closed cycle in $H_1(\Sigma,\ZZ)$, we combine such a saddle connection with its reflected mirror image on the other sheet of $\bar\Sigma\to \bar C$. Vectors correspond to closed geodesics which avoid both branch points and poles of $\lambda$; again, to get a cycle odd under the deck transformation we must combine this with its reflected mirror image. The moduli space of a closed geodesic is generically an interval, in accordance with the analogous observation for the higher-dimensional cycles corresponding to these geodesics. Note that we can recast the geodesics of interest as saddle connections or closed geodesics on $C$ -- the metric \eqref{eq:metric} is invariant under $\lambda\mapsto -\lambda$, and so it is well-defined on $C$. However, note that we are only interested in certain closed loops on $C$ -- namely, those which lift to closed loops on $\Sigma$.

\subsection{Low-energy physics on the Coulomb branch} \label{sec:lowNRG}

Next, we turn to the description of the pure $U(1)^r$ gauge theories that describe our UV theories at low energies in a neighborhood of the Coulomb branch. In $\N=1$ language, they are non-linear sigma models coupled to abelian gauge theories whose coupling constants depend on the scalars in the chiral multiplets. They are completely specified by the holomorphic prepotential $F(a)$, where $a$ refers to the scalars in the chiral multiplets $A^i$, $i=1,\ldots,r$. In $\N=1$ superspace, the Lagrangian is
\be \frac{1}{4\pi} \Imag\brackets{ \int d^4\theta\, A^{i\dagger} \frac{\partial F(A)}{\partial A^i} - \frac{1}{2} \int d^2\theta\, \frac{\partial^2 F(A)}{\partial A^i \partial A^j} W^{i\alpha} W^j_\alpha} \ .\ee
In particular,
\be \tau_{ij}(a)=\frac{\partial^2 F(a)}{\partial a^i \partial a^j} \label{eq:lowCouplings} \ee
both play the role of the $U(1)^r$ coupling constants and determine the K\"ahler metric
\be \Imag{\tau_{ij} da^i d\bar a^j} \ee
of the Coulomb branch. The electric and magnetic central charges of the $\N=2$ algebra are respectively given by
\be Z_{\beta^i} = a^i \label{eq:elZ} \ee
and
\be Z_{\alpha_i} = \frac{\partial F}{\partial a^i} \ ; \label{eq:magZ} \ee
here, $\{\alpha_i\}$ and $\{\beta^i\}$ are bases for the electric and magnetic charge lattices which satisfy
\be \avg{\alpha_i,\alpha_j} = \avg{\beta^i,\beta^j}=0\ ,\quad \avg{\alpha_i,\beta^j} = \delta^j_i \ .\ee
(There are also flavor central charges, but these are the same as the mass parameters of our theory, which are constant on $\B$.) Of course, choosing such a basis requires a choice of Lagrangian decomposition $\Gamma\cong \Gamma_m\oplus \Gamma_e$, and this choice of duality frame famously cannot be extended to all of $\B$ \cite{sw:theory1,sw}. So, a global description requires patching together different duality frames via $Sp(2r,\ZZ)\ltimes \ZZ^{2r}$ transformations of the form
\be \column{\gamma^g}{\gamma^f} \mapsto \column{N\gamma^g + M \gamma^f}{\gamma^f} \ ; \label{eq:duality} \ee
here, $\gamma^g\in \Gamma_u$ and $\gamma^f\in \Gamma_{\rm flavor}$, while $N\in Sp(2r,\ZZ)$ and $M\in \Hom(\Gamma_{\rm flavor},\Gamma_u)$. \eqref{eq:duality} makes it clear that the decomposition $\hat\Gamma_u \cong \Gamma_u\oplus \Gamma_{\rm flavor}$ only exists locally.

This IR data can be geometrically encoded in a fibration $\M$ of abelian varieties (special $2r$-tori) over $\B$ \cite{w:integrable}. The periods of the fiber $\T_u$ over $u$ are given by $\tau_{ij}$. To see $\M$ arise from physics, one may compactify the 4d field theory on a circle $S^1_R$ of radius $R$ \cite{sw:3d}. The moduli space of the resulting theory is hyper-K\"ahler, and in a canonical complex structure is precisely $\M$. (This canonical complex structure is independent of $R$. In this complex structure, $\M$ is the total space of a special Lagrangian fibration.) Indeed, $\M$ is the moduli space of solutions to Hitchin's equations with gauge group $SU(2)$ on $C$ (with prescribed singularities at the punctures). This follows by first compactifying our M5-branes on $S^1_R$ to obtain D4-branes wrapping $C$, since the vacua of this twisted 5d $SU(2)$ gauge theory on $C$ are defined by solutions of Hitchin's equations \cite{w:MGaugeSol,kapustin:to3d,kapustin:nonLag,kapustin:periodicMono,kapustin:singularPeriodicMono,GMN:classS}. (The topological twist of our 6d theory on $C\times S^1_R$ renders its vacuum structure insensitive to the conformal scale of $C$, so we may take it to be much larger than $R$ and still learn about the theory of interest, which exists in the opposite limit.)

We have now given two geometric characterizations of the (gauge) central charges $Z_\gamma$: they are related via \eqref{eq:lowCouplings}, \eqref{eq:elZ}, and \eqref{eq:magZ} to the periods of $\T_u$, and they are related via \eqref{eq:ZSW} to periods of $\lambda$ on $\bar\Sigma_u$. To relate these pictures, we note that\footnote{Here, $\partial_j\lambda$ is really the Lie derivative.}
\be \tau_{ij} = \partial_j Z_{\alpha_i} = \int_{\alpha_i}\partial_j\lambda = \int_{\alpha_i} \brackets{\iota_{\xi_j}\omega + d\parens{\iota_{\xi_j} \lambda}} = \int_{\alpha_i} \iota_{\xi_j}\omega \ .\ee
Here, $\xi_j$ is a section of the normal bundle of $\Sigma_u$ which indicates how the latter deforms as we vary $a^j$ (and in particular, which vanishes at the punctures of $\Sigma_u$). Furthermore, $\iota_{\xi_j} \omega$ is \emph{holomorphic} (and not merely meromorphic \cite{sw:theory1}) on $\bar\Sigma_u$; for, $\omega$ becomes ill-defined near punctures where $x$ diverges, but $\xi_j$ vanishes sufficiently rapidly to compensate for this. Indeed, the distinction between moduli and other coefficients in $\varphi$ is that the former correspond to deformations $\xi_j$ with $\iota_{\xi_j} \omega$ holomorphic \cite{mikhailov:surface}.

Thus, we have a set of $r$ holomorphic one-forms $\iota_{\xi_j} \omega$ on $\bar\Sigma_u$ whose periods associated to the cycles $\alpha_i$ are precisely $\tau_{ij}$. One might therefore expect that $\T_u=J(\bar\Sigma_u)$, the Jacobian $H^0(\bar\Sigma_u,\Omega^{1,0})^*/H_1(\bar\Sigma_u,\ZZ)$ of $\bar\Sigma_u$. (Here, the image of $\gamma\in H_1(\bar\Sigma_u,\ZZ)$ in $H^0(\bar\Sigma_u,\Omega^{1,0})^*$ is given by
\be \kappa\mapsto \int_\gamma \kappa \ , \ee
where $\kappa\in H^0(\bar\Sigma_u,\Omega^{1,0})$.) However, this cannot be right, as $\bar\Sigma_u$ is generically not of genus $r$, and so we must have picked out special 1-cycles and 1-forms. Indeed, we know this to be the case -- we restricted to the subset of 1-cycles which are odd under the deck transformation. The resulting subset of the Jacobian is the Prym variety
\be \Pp_u = \ker\parens{Nm: J(\bar\Sigma_u)\to J(\bar C)} \ , \label{eq:prym} \ee
where $Nm$ is the norm map
\be \parens{T^*\bar\Sigma_u \ni \kappa \mapsto f(\kappa)} \mapsto \parens{T^*\bar C\ni \tilde \kappa \mapsto f(p^*\tilde\kappa)} \ ,\ee
and $p:\bar\Sigma_u\to \bar C$ is the covering map.\footnote{In 4d, only the periods of $\T_u$ -- and not the actual points of $\T_u$ -- are important, and so this identification of $\T_u$ with the Prym variety is adequate. However, in actuality the fibers of the Hitchin fibration are torsors for the Prym variety \cite{w:integrable}.}

It is not uncommon -- especially in older literature -- to find `Seiberg-Witten curves' which are quotients of our $\Sigma$'s by symmetries under which $\lambda$ is invariant \cite{donagi:integrable,warner:geodesics,caceres:prym}. These quotients yield curves -- often with smaller genera or with fewer singularities -- whose Jacobian coincides with the above Prym variety. We will sometimes count geodesics on this curve instead, as the physics is nearly equivalent (the spectrum of BPS states may differ in a simple way, as the quotient curve can have new cycles that are halves of the original ones) and the math is simpler.

\subsection{Relationships between $(\Sigma,\lambda)$ and $(C,\phi)$} \label{sec:rels}

We comment in this section on relationships between UV and IR data. First, we relate the zeroes and poles of $\lambda$ and $\phi$. It is clear that poles (zeroes, resp.) of even order $N$ in $\phi$ correlate with poles (zeroes) of order $N/2$ in $\lambda$. Now, consider a singularity $\phi \sim \Lambda^2\, dz^2/z^N$ where $N$ is odd. In this case, $z$ is not a good coordinate on $\Sigma$ near $z=0$, since $\lambda \sim \pm \Lambda \, dz/z^{N/2}$ does not look like a meromorphic differential. So, we define $\zeta=z^{1/2}$ and employ the substitution $dz=2\zeta d\zeta$. (In contrast, if $N$ is even then $z$ is a good coordinate, in which case $\zeta$ is multivalued.) We then have
\be \lambda \sim \pm \frac{2\Lambda}{\zeta^{N-1}} \parens{1 + a_1 \zeta^2 + a_2 \zeta^4 + \ldots} d\zeta \ . \label{eq:noMass} \ee
This reasoning demonstrates the following important facts. First, poles (zeroes, resp.) of odd order $N$ in $\phi$ translate to poles (zeroes) of order $N-1$ ($N+1$, resp.) in $\lambda$. In particular, a quadratic differential $\phi$ with at most simple poles translates to a holomorphic differential $\lambda$ on $\bar\Sigma$. Second, as the good local coordinate $\zeta=z^{1/2}$ winds once around an odd order pole (in $\phi$), $z$ winds twice. That is, such a pole is a branch point, as can be verified by noting that $\lambda\mapsto -\lambda$ as $\zeta\mapsto -\zeta$. It corresponds to a smooth point of $\bar\Sigma$ if and only if the pole in $\phi$ is simple. (The only other branch points are the zeroes of $\phi$.)

Next, we explain the following useful reasoning that relates the genus $g'$ of $\Sigma$ to the genus $g$ of $C$. The Riemann-Hurwitz theorem relates the Euler characteristics of $C$ and $\Sigma$:
\be 2-2g' = 2(2-2g) - b \ ,\ee
where $b$ is the number of branch points. In turn, $b=z+o$, where $z$ is the number of zeroes of $\phi$ (which we always take to be simple) and $o$ is the number of odd poles (counted without multiplicity). The degree of the canonical bundle of $C$ is $2g-2$, so $z-p=4g-4$, where $p$ is the number of poles (counted with multiplicity). Combining these results, we have
\be g' = 4g-3+\frac{p+o}{2} \ . \label{eq:gP} \ee
Note that this is always an integer, since each odd pole makes an odd contribution to $p$.

\subsection{From M5-brane geometries to field theories}
\label{sec:examples}

We close this review by describing -- somewhat telegraphically -- the rules associating concrete field theories to M5-branes wrapping Riemann surfaces.
Most of these rules can be inferred by engineering Lagrangian quiver gauge theories with D4-, D6-, and NS5-branes in type IIA string theory and then taking an M-theory limit \cite{w:MGaugeSol,gaiotto:classS,GMN:classS}, or by applying mirror symmetry to our type IIB threefold compactifications \cite{vafa:geoEng,vafa:sol}, but for conciseness we will simply state the final results. We begin by introducing three classes of superconformal building blocks: the so-called trinion $T$ \cite{gaiotto:classS}, also known as 4 free hypermultiplets, as well as two classes of Argyres-Douglas theories \cite{GMN:classS,vafa:RTwist,xie:AD}, $X_N$, $N\ge 3$, and $Y_N$, $N\ge 6$. $T$ is constructed by taking $C$ to be a thrice-punctured sphere with simple poles at the punctures. $X_N$ (also known as $(A_1,D_{N-2})$) is associated to genus 0 curves $C$ with two punctures, one simple and one of order $N$; in particular, $X_3$ is an empty theory and $X_4$ is two free hypermultiplets, while the rest of this family is comprised of non-Lagrangian Argyres-Douglas CFTs \cite{cecotti:moreN2}. Finally, $Y_N$ (also known as $(A_1,A_{N-5})$) is associated to a genus 0 curve $C$ with a single irregular puncture of order $N$; $Y_6$ is a single free hypermultiplet, and $Y_7$ is the original Argyres-Douglas CFT associated to both pure $SU(3)$ \cite{argyresDouglas} and $SU(2)$ with one flavor \cite{sw:argyresDouglas,w:integrable}. $Y_8=X_5$ is the Argyres-Douglas CFT associated to $SU(2)$ with two flavors, while $X_6$ is associated to $SU(2)$ with three flavors. More generally, $Y_{N+4}$ is associated to pure $SU(N)$ and $X_{N+2}$ is associated to pure $SO(2N)$ \cite{eguchi:ad}.

Punctures have associated flavor symmetries, which are important for both gauging and turning on mass parameters. Regular punctures have associated $SU(2)$ symmetries; for example, $T$ can be thought of as a half-hyper (or chiral) in the trifundamental of $SU(2)^3$. (The full global symmetry of $T$ is larger, but as will soon become apparent it is convenient to focus on $SU(2)$ subgroups.) In particular, it is convenient to think of the empty theory $X_3$ as having $SU(2)$ flavor symmetry, as we will then be able to gauge it (although of course the empty theory does not possess a mass parameter, which is generally associated to flavor symmetries). Furthermore, irregular singularities provide an additional flavor symmetry if (and only if) the order of the pole is even. In particular, $X_4$ can be thought of as a half-hyper in the bifundamental of $SU(2)^2$.

Mass parameters are coefficients of poles of $\phi$ which yield residues of $\lambda$. This fact allows us to understand the results of the last paragraph. Consider an irregular singularity in $\phi$ of the form
\be \phi \sim \parens{\frac{\Lambda^2}{z^N} + \ldots + \frac{2m\Lambda}{z^{\frac{N}{2}+1}} + \ldots} dz^2 \ , \label{eq:phiMass} \ee
where $N>2$ is even. \eqref{eq:swPhi} implies that
\be \lambda \sim \pm \parens{\frac{\Lambda}{z^{N/2}} + \ldots + \frac{m}{z} + \ldots } dz \ , \label{eq:lambdaMass} \ee
and so $m$ is indeed a mass parameter. (This is a bit schematic, as other terms may contribute to the residue in \eqref{eq:lambdaMass}, but it suffices to demonstrate the existence of a mass parameter.) However, if $N>1$ is odd, then \eqref{eq:noMass} makes it clear that $\lambda$ cannot have a residue, so we simply cannot add a mass.  The would-be flavor charge associated to winding around such a pole is even under the deck transformation (if this cycle is even homologically non-trivial in the first place), since the pole is a branch point. The non-existence of this flavor charge is of course closely related with the non-existence of a mass parameter.

Next, we focus on simple singularities. To add a mass parameter, one upgrades the pole to a double pole of the form
\be \phi \sim \parens{\frac{m^2}{z^2} + \frac{\xi}{z} + \ldots} dz^2 \ , \ee
so that
\be \lambda \sim \pm \parens{\frac{m}{z} + \frac{\xi}{2m} + \ldots} dz \ . \ee
Finally, given any two simple punctures we can gauge the diagonal subgroup of the associated flavor symmetry by replacing the two punctures by a cylinder. Conversely, pinching off a cylinder so that it separates and we add two regular punctures corresponds to taking the gauge coupling of the $SU(2)$ gauge group associated to the cylinder to vanish.

These rules allow one to construct a wide variety of generalized $SU(2)$ quiver gauge theories, possibly coupled to Argyres-Douglas CFTs. The superconformal theories are the non-Argyres-Douglas (non-AD, henceforth) theories with only $T$ building blocks and vanishing mass parameters -- that is, theories with at most simple punctures -- and the Argyres-Douglas theories $X_N,Y_N$ (again, with vanishing mass parameters). In particular, theories including gauged $X_N$ building blocks are asymptotically free, thanks to the contribution of the gauged $X_N$ theory to the beta function of the corresponding $SU(2)$ gauge coupling \cite{s:scaling,xie:AD}. In contrast, the coupling constants of $SU(2)$ gauge groups associated to cylinders connecting (possibly coincident) trinions are always exactly marginal. These considerations make it clear when a cylinder contributes a new dynamical scale and when it contributes a marginal coupling constant. In particular, non-AD SCFTs come in moduli spaces of the form $\M_{g,n}$, where $g$ is the genus of $C$ and $n$ is the number of simple punctures. Since these moduli spaces have multiple degeneration limits associated to weakly-coupled gauge theories, the mapping class group gives us many dualities involving (generically strongly-coupled) gauge theories. More generally (even for non-conformal theories), the geometry of the UV curve gives us a rich web of dualities. Finally, we note that for non-AD theories (with vanishing mass parameters), the Coulomb branch is precisely \eqref{eq:affine}, and not an affine subspace thereof.

\begin{figure}
\begin{center}
\includegraphics[scale=1.2]{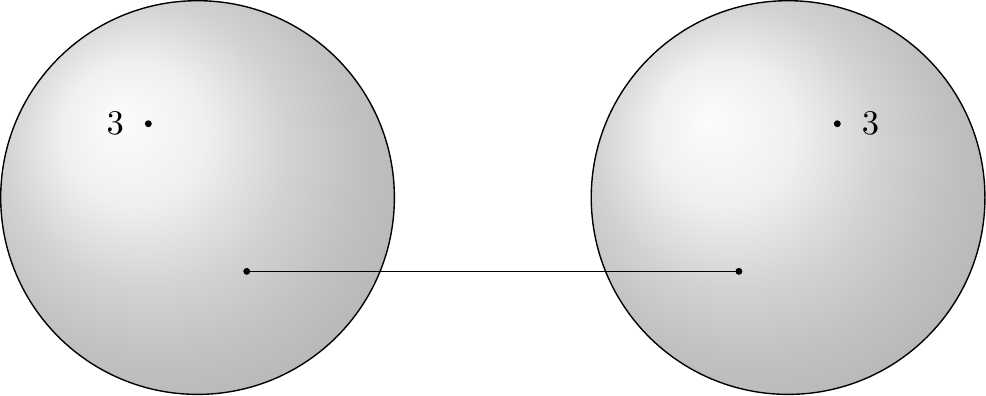}
\caption{The UV curve of $SU(2)$ SYM with $N_f=0$. Each sphere represents an $X_3$ factor, and the solid line between the two simple punctures is the cylinder connecting them.}
\label{fig:su2c}
\end{center}
\end{figure}

\begin{figure}
\begin{center}
\begin{subfigure}[b]{.45\textwidth}
\begin{center}
\includegraphics[scale=2]{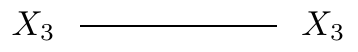}
\caption{}
\label{fig:su2schematic}
\end{center}
\end{subfigure}
~
\begin{subfigure}[b]{.45\textwidth}
\begin{center}
\includegraphics[scale=2]{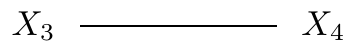}
\caption{}
\end{center}
\end{subfigure}\\\vspace{1cm}

\begin{subfigure}[b]{.45\textwidth}
\begin{center}
\includegraphics[scale=2]{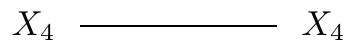}
\caption{}
\end{center}

\end{subfigure}
~
\begin{subfigure}[b]{.45\textwidth}
\begin{center}
\includegraphics[scale=2]{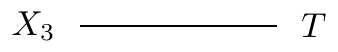}
\caption{}
\end{center}

\end{subfigure}\\ \vspace{1cm}

\begin{subfigure}[b]{.45\textwidth}
\begin{center}
\includegraphics[scale=2]{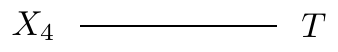}
\caption{}
\end{center}

\end{subfigure}
~
\begin{subfigure}[b]{.45\textwidth}
\begin{center}
\includegraphics[scale=2]{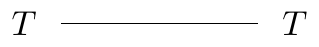}
\caption{}
\end{center}

\end{subfigure}\\\vspace{1cm}

\begin{subfigure}[b]{.45\textwidth}
\begin{center}
\includegraphics[scale=2]{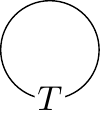}
\caption{}
\end{center}

\end{subfigure}

\caption{Schematic representations of the UV curves of (a) $SU(2)$ $N_f=0$, (b) $SU(2)$ $N_f=1$, [(c) and (d)] $SU(2)$ $N_f=2$, (e) $SU(2)$ $N_f=3$, (f) $SU(2)$ $N_f=4$, and (g) $\N=2^*$ $SU(2)$. Solid lines represent punctures of $C$ being connected by a cylinder, or more physically an $SU(2)$ gauge group.}
\label{fig:schematics}
\end{center}
\end{figure}

The simplest example is $SU(2)$ super Yang-Mills (SYM) with $N_f=0$. This arises from $C$ of genus $g=0$, with two irregular singularities of order 3. This follows from connecting two copies of $X_3$ with a cylinder. The Seiberg-Witten curve is $\lambda^2 = \parens{\frac{\Lambda^2}{z^3} + \frac{u}{z^2} + \frac{\Lambda^2}{z}}\, dz^2$. We have drawn the UV curve of this theory in Figure \ref{fig:su2c}, and a schematic of this construction is shown in Figure \ref{fig:su2schematic}.

Many familiar $SU(2)$ theories can be constructed from UV curves with a similar structure. Some examples are:
\begin{itemize}
\item $SU(2)$ $N_f=1$ has $g=0$ with two irregular singularities of order 3 and 4, respectively. This follows from connecting $X_3$ to $X_4$.
\item $SU(2)$ $N_f=2$ has $g=0$ with two irregular singularities of order 4. This follows from connecting two copies of $X_4$. Alternatively, we can have $g=0$ with two regular singularities and an irregular one of order 3, as follows from connecting $X_3$ to $T$. These two constructions are related \cite{GMN:classS} via the Hanany-Witten effect \cite{hananyWitten}.
\item $SU(2)$ $N_f=3$ has $g=0$ with an irregular singularity of order 4 and two regular singularities. This follows from connecting $T$ to $X_4$.
\item $SU(2)$ $N_f=4$ has $g=0$ with 4 regular singularities. This follows from connecting two copies of $T$. It is superconformal and has an exactly marginal coupling constant parametrized by $\M_{0,4}=\HH/SL(2,\ZZ)$.\footnote{$PSL(2,\CC)$ transformations can fix 3 marked points at $0,1,\infty$, so $\M_{0,4}$ consists of the points in $\CC\backslash\{0,1\}$. To such a point $\xi$, we can associate the elliptic curve $y^2 = z(z-1)(z-\xi)$. It follows that $\M_{0,4}\cong \M_{1,1}$.} The Seiberg-Witten curve is of the form
\be \lambda^2 = \parens{ \frac{P(z)}{(z-\xi)^2(z-1)^2 z^2} + \frac{u}{(z-1)(z-\xi)z}}\,dz^2 \ ,\ee
where $P(z)$ is a polynomial of degree at most 3 which is homogeneous of degree two in the mass parameters and independent of $u$. $\xi=\lambda(\tau)$ (where $\lambda$ here is the modular lambda function) gives a non-perturbative regularization scheme for defining the coupling constant $\tau$. With this definition, $\tau$ is the modular parameter of the Seiberg-Witten curve when $P(z)\equiv 0$, or equivalently when $|u|\to \infty$.
\item $SU(2)$ $\N=2^*$ (that is, $\N=4$ with a mass for the adjoint hypermultiplet) can more or less be engineered by choosing $C$ to be of genus $g=1$ with an order 2 pole. This follows from connecting two punctures of $T$: the ${\bf 2}\otimes{\bf 2}={\bf 3}\oplus{\bf 1}$ representation of $SU(2)$ yields an adjoint hypermultiplet and a neutral hypermultiplet that decouples. This theory is not superconformal, but it nevertheless has an exactly marginal gauge coupling which parametrizes $\M_{1,1}=\HH/SL(2,\ZZ)$, the moduli space of elliptic curves. In particular, it has S-duality. The Seiberg-Witten curve is $\lambda^2 = (m^2 \wp(z|\tau) +u) \, dz^2$, where $z\sim z+1\sim z+\tau$ and $\wp(z|\tau)$ is the Weierstrass $\wp$ function which behaves as $1/z^2$ near $z=0$.
\item By turning off the mass parameter associated to the pole in the last example, we learn that $g=1$ with no punctures gives $\N=4$ $SU(2)$; $\Sigma$ is simply two parallel copies of $C$. (There is no quadratic differential on $C$ with a simple pole, so turning off the mass parameter completely eliminates the singularity of $\phi$ at $z=0$.)
\end{itemize}
We have drawn schematic constructions of the UV curves for these theories in Figure \ref{fig:schematics}. 

\begin{figure}
\begin{center}
\begin{subfigure}[b]{.45\textwidth}
\begin{center}
\includegraphics[scale=1]{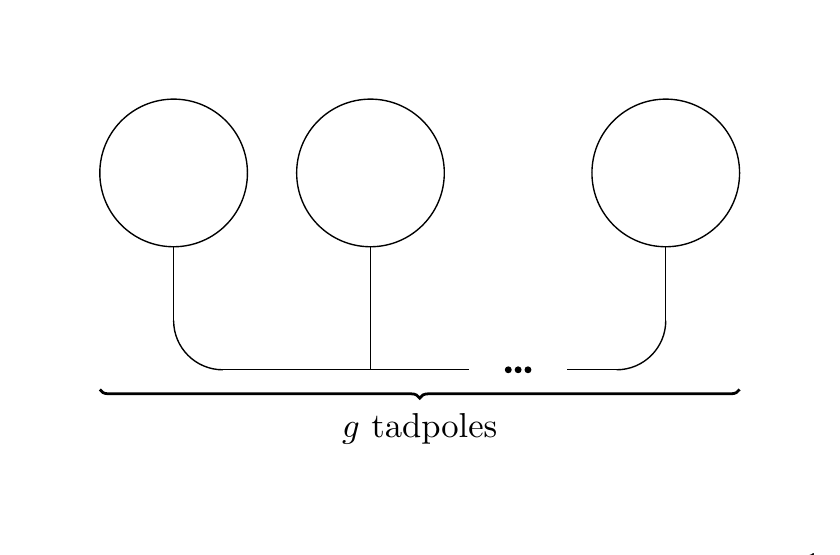}
\caption{}
\label{fig:tadpoles}
\end{center}
\end{subfigure}
~
\begin{subfigure}[b]{.45\textwidth}
\begin{center}
\includegraphics[scale=1.5]{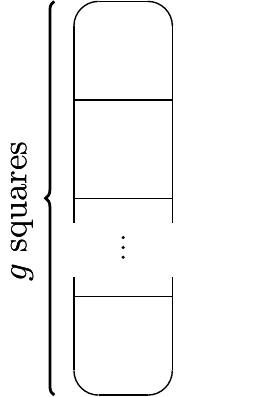}
\caption{}
\label{fig:boxes}
\end{center}
\end{subfigure}
\caption{Trivalent graphs representing two Lagrangian descriptions of theories with purely holomorphic $\phi$. Vertices represent copies of the trinion $T$ and edges represent cylinders connecting them, i.e. $SU(2)$ gauge groups. We have rounded the outer edges of the graphs to emphasize that these are not vertices.}

\end{center}
\end{figure}

For our final example, we provide two different (generically strongly coupled) Lagrangian descriptions of the SCFTs associated to $C$ with $g>1$ and no punctures, generalizing \S9.5.4 of \cite{tachikawa:N2}. To do so, we study different weakly coupled degeneration limits, where we have a number of copies of $T$ coupled via thin cylinders. We can describe such a limit with a trivalent graph whose vertices correspond to a copy of $T$ and whose edges correspond to these cylinders. We note that the absence of any poles means that we cannot turn on mass parameters for any of the matter in these gauge theories.
\begin{itemize}
\item One such graph follows by attaching $g$ tadpoles on top of a horizontal line segment, as shown in Figure \ref{fig:tadpoles}. This yields a $\prod_{i=1}^g SU(2)_{a,i} \times \prod_{i=2}^{g-1} SU(2)_{b,i} \times \prod_{i=1}^{g-1} SU(2)_{c,i}$ gauge theory, where the $a$ factors are associated to the loops in the tadpoles, the $b$ lines connect the $i\not=1,g$ tadpoles to the base, and $c$ lines connect horizontally to the vertices in the base. For $i\not=1,g$, we have a half-hyper in the $({\mathbf 3}\oplus{\mathbf 1})\otimes{\mathbf 2}$ of $SU(2)_{a,i}\times SU(2)_{b,i}$ and another in the ${\mathbf 2}\otimes{\mathbf 2}\otimes{\mathbf 2}$ of $SU(2)_{b,i}\times SU(2)_{c,i-1}\times SU(2)_{c,i}$; finally, we have half-hypers in the $({\mathbf 3}\oplus{\mathbf 1})\otimes{\mathbf 2}$ of $SU(2)_{a,1}\times SU(2)_{c,1}$ and $SU(2)_{a,g}\times SU(2)_{c,g-1}$. 

\item Another limit corresponds to a graph comprised of $g$ squares stacked on top of each other, as shown in Figure \ref{fig:boxes}. This has $\prod_{i=0}^g SU(2)_{a,i} \times \prod_{i=2}^{g-1} SU(2)_{\ell,i} \times \prod_{i=2}^{g-1} SU(2)_{r,i}$ gauge group, associated to the horizontal, left, and right lines, respectively.

First, suppose that $g=2$. Then, we have two half-hypers in the ${\mathbf 2}\otimes {\mathbf 2}\otimes {\mathbf 2}$ representation of $SU(2)_{a,0}\times SU(2)_{a,1}\times SU(2)_{a,2}$. In particular, there is an $SU(2)$ flavor symmetry rotating the two half-hypers to which there is no associated mass parameter in the class S construction.

Now, consider $g>2$. For $i=2,\ldots,g-2$, we have half-hypers in the ${\mathbf 2}\otimes {\mathbf 2}\otimes {\mathbf 2}$ representation of $SU(2)_{a,i}\times SU(2)_{\ell,i}\times SU(2)_{\ell,i+1}$ and of $SU(2)_{a,i}\times SU(2)_{r,i}\times SU(2)_{r,i+1}$. We also have half-hypers in the ${\mathbf 2}\otimes {\mathbf 2}\otimes {\mathbf 2}$ representation of $SU(2)_{a,0}\times SU(2)_{a,1}\times SU(2)_{\ell,2}$, $SU(2)_{a,0}\times SU(2)_{a,1}\times SU(2)_{r,2}$, $SU(2)_{a,g-1}\times SU(2)_{a,g}\times SU(2)_{\ell,g-1}$, and $SU(2)_{a,g-1}\times SU(2)_{a,g}\times SU(2)_{r,g-1}$.
\end{itemize}
Note that the number of gauge groups ($3g-3$) is duality invariant, since this is the dimension $r$ of the Coulomb branch. It equals the dimension of the moduli space $\M_{g,0}$, since each coupling constant is associated to a gauge group.

\section{Geodesics on Flat Surfaces}
\label{sec:geo}

The basic objects in the class S construction are a quadratic differential $\phi$, defined on a Riemann surface $C$ of genus $g$, and a one-form -- or ``abelian differential" -- $\l$ defined on another Riemann surface $\Sigma$ of genus $g'$. We call a surface equipped with an abelian differential a translation surface, and a surface with a quadratic differential a half-translation surface.\footnote{Throughout this section, we will treat the singular points of a surface as part of the surface, in contrast to earlier sections.} Quadratic and abelian differentials are examples of a broader type of objects known as $k$-differentials, i.e. tensors $f_k$ of the form \eq{f_k = f(z)\, dz^k} defined on Riemann surfaces; abelian differentials have $k=1$, and quadratic differentials have $k=2$. We call $f_k$ holomorphic if $f(z)$ has no poles of order $k$ or greater, and meromorphic otherwise. The name ``flat surface" is common for surfaces equipped with a $k$-differential.  To see why, note that
\be ds^2 = \abs{f(z)}^{2/k} dz d\bar z \ee
defines a flat metric (possibly with conical singularities).

In general, a $k$-differential will have zeroes and poles; we will label the orders of the $m$ zeroes of $f_k$ by $a_i$, $i = 1,\ldots,m$, and similarly say $f_k$ has $n$ poles of order $b_j$, $j = 1,\ldots,n$. We also allow surfaces to have marked points, which we consider as zeroes of order zero of their associated $k$-differential; we will motivate the inclusion of such marked points shortly. The Riemann-Roch theorem implies that \eq{\sum_{i}a_i - \sum_jb_j = k\left(2g-2\right),\label{eq:RRk}} where $g$ is the genus of the Riemann surface on which $f_k$ is defined. The moduli space $\mathcal{K}_g$ of biholomorphism equivalence classes of pairs $(C,f_k)$, where $f_k$ is a $k$-differential on a genus $g$ surface $C$, naturally decomposes as a disjoint union   of ``strata" $\mathcal{K}_g(a_1,\cdots,a_m,-b_1,\cdots,-b_n)$ within which the differential has a fixed singularity structure.  We will frequently denote the singularity structure $\{a_1,\cdots,a_m,-b_1,\cdots,-b_n\}$ as $\b$ for brevity. If we restrict ourselves to have no poles or marked points, there are only finitely many strata of $k$-differentials of fixed genus, but if we allow poles of any order then there exist infinitely many strata. We refer to poles, marked points, and zeroes as singular points.

Although this is not a standard definition, we will allow strata of $k$-differentials to include tensors that are globally $k$-th powers of an abelian differential, i.e. $k$-differentials of the form $\l^k$ for an abelian differential $\l$. If $\l$ is in $\Hh(a_1,\cdots,a_m,-b_1,\cdots,-b_n)$, then $\l^k$ is in $\mathcal{K}(ka_1,\cdots,ka_m,-kb_1,\cdots,-kb_m)$. Therefore, since $\phi$ generically has simple zeroes, this subtlety will mostly be unimportant, but it will be essential for one of our main examples.

For our purposes, we will only need to consider abelian and quadratic differentials, so we will immediately specialize to those cases, for which we will introduce slightly different notation. In particular, we will write strata of abelian differentials as $\Hh(\b')$, and on the other hand we will write strata of quadratic differentials as $\Q(\b)$. We have introduced primes next to the orders of singularities of quadratic differentials to avoid confusion between the two cases; we will similarly write $g'$ for the genus of a Riemann surface equipped with an abelian differential, and $g$ for the genus of a surface with a quadratic differential. Specializing \eqref{eq:RRk} to $k=1$ and $k=2$, we have \begin{subequations} \eq{\sum_{i=1}^m a_i' - \sum_{j=1}^n b_j' = 2g'-2 \label{eq:RRmero}} for abelian differentials and \eq{\sum_{i=1}^m a_i - \sum_{j=1}^n b_j = 4g-4,\label{eq:RRquad}} for quadratic differentials. \end{subequations} 

In \S \ref{sec:max} we explained the relationship between the quadratic differential $\phi$ on the UV curve $C$ and the differential $\l$ on $\Sigma$. In the language of dynamical systems, this relationship is known as the ``canonical double cover" $\frakC: \Q(\beta)\to \Hh(\beta')$. Starting with a pair $(\phi,C)$, $\fracC$ produces a double cover $\Sigma \subset T^*C$ of $C$, branched only over singular points of $\phi$, equipped with an abelian differential $\l$ which squares to the pullback of $\phi$ under the covering map. Physically, this is the Seiberg-Witten curve -- defined by the equation $\lambda^2 = \phi$ -- and its associated differential. 

Away from its zeroes and poles, any quadratic differential can be written as the square of an abelian differential \cite{chen:kDiff,strebel:quad}. However, as explained in \S\ref{sec:rels} the relationship between $(C,\phi)$ and $(\Sigma,\lambda)$ at the singular points is more complicated. The stratum $\Hh(\b')\ni(\Sigma,\l)$ is obtained from the stratum $\Q(\b)\ni(C,\phi)$ as follows: for a singular point of $\phi$ of even order $2n$, $\l$ has two singular points of order $n$, and conversely for a singular point of order $2n-1$, we obtain only one point of order $2n$. In particular, $\frakC$ takes a simple pole of $\phi$ to a marked, but otherwise regular, point of $\l$. For instance, $\fracC$ takes the stratum $\Q(-1^4)$ to the stratum $\Hh(0^4)$ of holomorphic abelian differentials on the torus with four marked points, and the stratum $\Q(1^2,-3^2)$ to the stratum $\Hh(2^2,-2^2)$ of meromorphic abelian differentials. 

From these examples, it is clear that the genera of the Riemann surfaces $C$ and $\Sigma$ will in general not be equal. However, there is at least one important counterexample: the stratum $\Q(\O)$ of holomorphic quadratic differentials on a torus gets mapped to the stratum $\Hh(\O)$, which again consists of differentials defined on a torus. Importantly, this example only makes sense given our nonstandard definition of $\Q(\b)$:  $\Q(\O)$ is empty unless we include squares of abelian differentials \cite{masur:prescribed}. This is the only example we will consider where this subtlety will be important. Whenever we consider a quadratic differential $\phi$ that is globally a square of an abelian differential, as is the case for all differentials in $\Q(\O)$, the associated double cover is somewhat degenerate: it is simply two disjoint copies of the original surface. 

As detailed in \S\ref{sec:max}, both quadratic and abelian differentials induce metrics on the associated Riemann surfaces. These metrics are flat away from the zeroes and poles of the differential, where they develop conical singularities. These metrics are central objects of study in Teichmuller theory and related areas of dynamical systems. A natural course of study is to investigate the geodesics of these metrics associated to $k$-differentials. The two most interesting classes of geodesics are saddle connections, which run between two singular points, which if poles have order strictly less than $k$, and cross no other singular points; and closed geodesics, which we always take to avoid the conical singularities of $f_k$. In particular, saddle connections for quadratic differentials can terminate at simple poles of $\phi$, which get turned into marked points of $\l$; this is why we explicitly allow saddle connections to end on marked points.

The main point of this paper is that, in light of the relationship between BPS states and geodesics outlined above, we can use techniques and results from dynamical systems to study the BPS spectra of class S theories of $A_1$ type. In this section, we will outline the tools that we have at our disposal, and in particular we will emphasize known asymptotics of counts of saddle connections for abelian and quadratic differentials. First, in \S \ref{sec:holo} we will discuss classical results for counts of geodesics associated to holomorphic differentials, before moving on in \S \ref{sec:mero} to more recent progress on counting geodesics for meromorphic differentials. In our discussion of holomorphic differentials, we will mostly follow the excellent reviews in \cite{zorich:flat,eskin:notes}. On the other hand, for meromorphic differentials, our understanding is quite incomplete, and foundational new results are still being proven.

\subsection{Counting Geodesics for Holomorphic Differentials}
\label{sec:holo}
In this section, we will discuss geodesics on Riemann surfaces equipped with a holomorphic quadratic or abelian differential.

One of the key tools of dynamical systems is the action of $\operatorname{GL}(2,\RR)$ on strata of holomorphic abelian or quadratic differentials. The origin of this action is quite transparent. Excluding singular points, a flat surface with a $k$-differential may be covered by coordinate charts whose transition functions take the form $z\mapsto \zeta_k z + c$, where $\zeta_k$ is a $k$-th root of unity (which can vary between the different transition functions). Then, thinking of $z$ as a vector in $\RR^2$ provides a natural linear action of $\operatorname{GL}(2,\RR)$ on coordinates which, when $k$ is 1 or 2, preserves the above form of the transition functions. ($-1$ commutes with all of $\operatorname{GL}(2,\RR)$; this is not the case for more general rotations $\zeta_k\in \operatorname{SO}(2,\RR)$.) This becomes an action on strata of flat surfaces, as opposed to merely a diffeomorphism, when we specify the action on the associated differential: $dz^k\mapsto dz^k$. The study of this action, and its associated flow on moduli space, is a central hallmark of the study of dynamical systems \cite{zorich:flat}.

Thanks to this $\operatorname{GL}(2,\RR)$ action, it turns out that as long as $\phi$ is holomorphic, counting problems on a half-translation surface $(C,\phi)$ and its double cover $(\Sigma,\l)$ are quite similar, so we will start off by considering quadratic differentials and then return to abelian differentials towards the end of this section. We therefore restrict ourselves to study strata of the form $\Q(a_1,\cdots,a_m)$, with no entries less than $-1$.

A quadratic differential $\phi$ defined on a Riemann surface $C$ defines a metric on $C$.  We would like to compute the numbers $N(L,C)$ and $N_{\text{closed}}(L,C)$ of saddle connections and closed geodesics of length less than $L$ on a surface $(C,\phi) \in\Q(\b)$. (Since closed geodesics have moduli, really we are counting families of parallel closed geodesics.) This can be usefully reformulated as follows. Fix some saddle connection of length $\ell$ that leaves one of its terminal zeroes with angle $\theta$; because away from the zeroes the metric is flat, this angle does not change along the geodesic. $(\ell,\theta)$ are the polar coordinates of a point in $\RR^2$. The set of all saddles on $S$ therefore gives a discrete subset $\Gamma(C)\subset\RR^2$ of the plane centered at the origin, and computing $N(L,C)$ amounts to asking how many points of $\Gamma(C)$ live within a circle of radius $L$ centered at the origin. Similarly, we could construct the analogous set $\Gamma_{\text{closed}}(C)$ for closed geodesics, and again ask how many of its points live in this circle.

For tori, $\Gamma_{\text{closed}}(C)$ is given by the primitive vectors of a lattice, and it follows that its points are roughly equidistributed in $\RR^2$. This was used, e.g., in \cite{mz:slagCounts} to determine the asymptotics $N_{\text{closed}}(L,C)\sim L^2$ for tori. Remarkably, similar quadratic asymptotics hold for any half-translation surface with a holomorphic differential, both for closed geodesics and saddle connections. In particular, Masur \cite{masur:geodesics1,masur:geodesics2} showed that, for any half-translation surface, there exist positive real constants $c_1,c_2\in\RR^+$ such that \eq{c_1 L^2 \le N_{\text{closed}}(L,C) \le N(L,C) \le c_2L^2 \label{eq:masurQuad}} for asymptotically large $L$. For simplicity, we will usually write this result as \eq{N(L,C)\sim{L}^2 \ .\label{eq:masursim}} 
This formulation emphasizes that the `entropy' \be S=\log\frac{dN}{dL} \sim \log L \ee is independent of $c_1,c_2$, at large $L$.

This result is extremely powerful. It applies for any holomorphic half-translation surface whatsoever, in any stratum. However, it has one drawback: if one is interested in the constants $c_1$ and $c_2$, then they must be computed separately for each surface. In particular, it does not allow us to relate the asymptotic number of geodesics of two surfaces in a stratum. One reason for this is trivial.\footnote{This intuition was provided, in a slightly different context, by D. Aulicino.} To any surface $(C,\phi) \in\Q(\b)$ we can associate a family of surfaces $\{s(C,\phi) : s\in\RR^+\}$, obtained from $(C,\phi)$ by simply rescaling $\phi\to{s}\phi$. We then have \eq{N(L,C) = N(s^{1/2}L,sC) \ ,\label{eq:Nscaling}} so each surface in this infinite family has a different spectrum of saddle connections.

We can get around this by defining a ``measuring stick" that allows us to compare lengths of geodesics in different surfaces. Consider the real codimension one sublocus $\Q_1(\b)\subset\Q(\b)$ of unit area surfaces in $\Q(\b)$. This amounts to picking one representative of each line $sC$ introduced above. The action of $\operatorname{GL}(2,\RR)$ on $\Q(\b)$ reduces to the action of $\operatorname{SL}(2,\RR)$ on $\Q_1(\b)$. It was proven by Masur \cite{masur:measure} and Veech \cite{veech:measure} that there exists an $\operatorname{SL}(2,\RR)$ invariant measure\footnote{More carefully, the measure is ergodic with respect to this action on each connected component of the stratum.} $\mu$ on each stratum $\Q_1(\b)$, and moreover that each stratum has finite volume with respect to this measure.

Averaging $N(L,C)$ over all surfaces in the unit area slice of a fixed connected component of a stratum implies that \cite{eskin:siegelveech}  \eq{\lim_{L\to\infty}\frac{N(L,C)}{\pi L^2} &= c \text{ for almost all }C\in\Q_1(\b) {\label{eq:SVquadSaddle}} \\ \lim_{L\to\infty}\frac{N_{\text{closed}}(L,C)}{\pi L^2} &= c_{\text{closed}}\text{ for almost all }C\in\Q_1(\b) {\label{eq:SVquadClosed}} \ .} Put another way, generic surfaces in any fixed (connected component of a) stratum have the same asymptotic number of geodesics! This should be thought of as somewhat similar to an index: the exact saddle connections or closed geodesics of course depend on exactly what surface in the stratum we are interested in, but, apart from finite codimension loci, the asymptotic count is constant. 

The constants $c,c_{\text{closed}}$ are known as ``Siegel-Veech" (SV) constants.\footnote{As mentioned in the preceding footnote, the measure is $\operatorname{SL}(2,\RR)$-invariant on each connected component, and therefore we obtain a different Siegel-Veech constant for each connected component. In the cases we will consider, there will only be one connected component, and we can ignore this subtlety.} Many Siegel-Veech constants are known. For instance, the Siegel-Veech constants for a torus\footnote{\eqref{eq:RRk} implies that holomorphic $k$-differentials on a torus have no zeroes. To define saddle connections on tori, we therefore have to mark two points $z_1\neq z_2\in{T}^2$ and count geodesics between those two points \cite{zorich:flat}.} are known to be (see, e.g., \cite{zorich:flat,mz:slagCounts}) \begin{subequations} \eq{c_{g=1} &= 1\\ c_{g=1,\text{ closed}} &= \frac{1}{\zeta(2)} = \frac{6}{\pi^2}\label{eq:SVtorus} \ .} \end{subequations} SV constants were related to volumes of connected components of strata in \cite{zorich:volumes}. Algorithms for computing these volumes were provided in \cite{eskin:volumes,eskin:volumes2}; \cite{zagier:largeG} provided an alternative algorithm that worked recursively in the genus of a surface. Volumes for several strata are computed in \cite{zorich:quad,goujard:constants}. Large genus asymptotics were conjectured in \cite{eskin:largeG} and proven in \cite{zagier:largeG0,sauvaget:volumes,aggarwal:largeG2,aggarwal:largeG,zagier:largeG}.

We now turn to holomorphic abelian differentials, and therefore restrict ourselves to strata of the form $\Hh(a_1',\cdots,a_m')$, with no negative entries. \eqref{eq:RRmero} now reads \eq{\sum_{i=1}^m a_i' = 2g'-2 \label{eq:RRholo}\ .} These strata are complex orbifolds of dimension \eq{\dim_\CC \Hh(a_1',\cdots,a_m') = 2g'-1+m \label{eq:holoabeldim}\ .} Aspects of the the topology of strata of holomorphic abelian differentials, and in particular the number of connected components of each stratum, were studied by Kontsevich and Zorich \cite{kontsevich:connectedComponents}. (They also studied the question of which $\Sigma$ arise from some $C$.) After a lengthy classification, they found that no stratum has more than three connected components.

We define the counting functions $N'(L,\Sigma)$ and $N'_{\text{closed}}(L,\Sigma)$ for abelian differentials in the obvious way: $N'(L,\Sigma)$ counts the number of saddle connections of length less than $L$ on a surface $\Sigma$ with abelian differential $\l$, and $N'_{\text{closed}}(L,\Sigma)$ counts the number of (families of) closed geodesics. Then for any surface $\Sigma$ equipped with an abelian differential there exist positive constants $c_1',c_2'>0$ such that \cite{masur:geodesics1,masur:geodesics2} \eq{c_1'L^2\le N'_{\text{closed}}(L,\Sigma) \le N'(L,\Sigma) \le c_2'L^2 \ .\label{eq:masur}} Similarly, if we restrict ourselves to the substratum $\chob\subset\Hh(\b')$ of unit area surfaces, there exist constants $c'$, $c'_{\text{closed}}$ such that \begin{subequations} \eq{\lim_{L\to\infty}\frac{N'(L,\Sigma)}{\pi L^2} &= c' \label{eq:SVabelSaddle}\\ \lim_{L\to\infty}\frac{N'_{\text{closed}}(L,\Sigma)}{\pi L^2} &= c'_{\text{closed}}\label{eq:SVabelClosed}} \end{subequations} for almost all $\Sigma\in\chob$.  

There has been recent progress \cite{eskin:billiards1} toward combining the best features of equations \eqref{eq:masur}--\eqref{eq:SVabelClosed}. It is now known that
\be \lim_{T\to\infty} \int_0^T N'_{\rm closed}(e^s,\Sigma) e^{-2s} ds = c'_{\rm closed} \ . \ee
(Similar results were proven in \cite{zorich:quad}.) This holds everywhere, not just almost everywhere, and nearly gives exact asymptotics as in \eqref{eq:SVabelClosed}. Additional work on extending `almost everywhere' to `everywhere' appears in \cite{dozier:all}.

Given how similar counts of geodesics are for quadratic and abelian differentials, it is natural to ask whether we should count on $C$ or on $\Sigma$. To see why it is preferable to count saddle connections on $C$, consider the two strata $\Q(1^5,-1^5)$ and $\Q(1^5,-1)$, which both have double covers in $\Hh(2^5)$. There is no obvious reason that surfaces in these two strata should have the same asymptotic counts of saddle connections (and indeed, their Siegel-Veech constants were computed in \cite{goujard:constants},\footnote{The constants listed there are not exactly the constants in \eqref{eq:SVquadSaddle}, but are directly proportional to them. Thus, the fact that the constants in the table are not equal implies that the constants we need are also not equal.} where they were shown to differ). On the other hand, naively applying \eqref{eq:SVabelSaddle} to the stratum $\Hh(2^5)$ seems to suggest that these strata should have the same asymptotic count. The resolution is that the images of quadratic strata in the abelian strata are generically of finite codimension, and therefore \eqref{eq:SVabelSaddle} tells us nothing about the growth rate of saddle connections on $\Sigma$. Thus, to obtain Siegel-Veech asymptotics, we should count saddle connections on $C$, rather than on $\Sigma$. We can of course apply \eqref{eq:masur} to the image of quadratic differentials in abelian strata, but this tells us nothing that was not already known from \eqref{eq:masurQuad}.

We now turn to vector multiplets. The same reasoning leads us to want to count geodesics on $C$, as opposed to $\Sigma$. However, as explained in \S\ref{sec:max}, we are only interested in those geodesics on $C$ which lift to closed geodesics on $\Sigma$. Fortunately, this lifting is an $SL(2,\RR)$-equivariant condition, and so there exists a variant of \eqref{eq:SVquadClosed} that holds for this counting problem.

\subsection{Counting Saddle Connections for Meromorphic Differentials}
\label{sec:mero}

The above discussion applies if and only if the UV curve $C$ has no punctures of order greater than one. For generic theories we will need to count geodesics for meromorphic differentials. Whereas dynamics of holomorphic differentials has been studied for decades, the corresponding study for meromorphic differentials is more recent. In particular, strata of meromorphic abelian differentials were first studied carefully in \cite{boissy:moduli}, where the topology of their moduli spaces was explored in detail. Saddle connections for meromorphic differentials were studied in \cite{aulicino:cbrank,tahar:poles}; we will summarize their results in this section. On the other hand, closed geodesics with meromorphic differentials have not been well-studied, and therefore we will not have anything to say. 

The works \cite{aulicino:cbrank,tahar:poles} focused on meromorphic abelian differentials, so for theories with higher-order punctures we will count curves on $\Sigma$ instead of on $C$. We will thus consider strata of the form $\Hh(a_1',\cdots,a_m',-b_1',\cdots,-b_n')$, where $n\ge1$. These strata are complex orbifolds of dimension \eq{\dim_\CC \Hh(a_1',\cdots,a_m',-b_1',\cdots,-b_n') = 2g'-2+m+n \label{eq:meroabeldim} \ ,} where $g'$ is given in terms of $a_i'$ and $b_j'$ via \eqref{eq:RRmero}. 

The presence of poles in a holomorphic differential radically changes the dynamics on the associated surface. On a surface without poles, generic geodesics will wind around the surface, but once poles are present most geodesics will terminate at a pole. To formalize this intuition, we introduce the set $\Theta(\Sigma)\subset{S}^1$ of saddle connection directions on a surface $(\Sigma,\l)$. Whereas for holomorphic $\l$ it can be shown that $\Theta(\Sigma)$ is a dense subset of the unit circle \cite{masur:dense}, for meromorphic $\l$ $\Theta(\Sigma)$ is closed in the unit circle \cite{GMN:classS,tahar:poles,aulicino:cbrank}. (This set is generally quite interesting. \cite{vorobets:geodesics1} shows that saddle connections have equidistributing phases for almost all translation surfaces without poles. Similar equidistribution results are studied in \cite{dozier:equi}. \cite{athreya:gaps} studies the asymptotic behavior of the smallest gap in this set, again for almost all translation surfaces, and characterizes the exceptional set of translation surfaces with different asymptotics. \cite{athreya:gapsL} gives the limiting gap distribution in a special case, and \cite{davis:L} showed that $\Theta$ in this special case has an interesting tree structure.)

The set $\Theta(\Sigma)$ will be essential for allowing us to count saddle connections on $\Sigma$, but before we count these saddle connections it is useful to obtain a geometric picture of a surface equipped with a meromorphic differential. In light of the above discussion, it is natural to ask where on a fixed surface $\Sigma$ saddle connections are allowed to live. The answer was provided by \cite{kontsevich:flatStable}, and elaborated on in \cite{tahar:poles}. 

To obtain the answer, it is helpful to define a nonstandard notion of convexity. We call a subset $S$ of $\Sigma$ convex if every geodesic between any two points of $S$ lies entirely within $S$, and as usual we define the convex hull $S$ of a subset $S'$ of $\Sigma$ to be the smallest closed convex set $S\supset S'$. We then define the core $\core(\Sigma)$ of $\Sigma$ to be the convex hull of the zeroes of $\l$; we will also frequently refer to the boundary $\dcore(\Sigma)$ and interior $\icore(\Sigma)\equiv\core(\Sigma)\setminus\dcore(\Sigma)$ of $\core(\Sigma)$. Amusingly, the ``geodesic horizon" used to study BPS geodesics in \cite{warner:geodesics} is exactly the boundary of the core of the Seiberg-Witten curve, even though that paper was written nearly twenty years before \cite{kontsevich:flatStable} introduced the core of a translation surface with poles to the math literature.

By definition, all saddle connections must live inside $\core(\Sigma)$. If a saddle connection left the core, then the core would not be convex, which would violate its definition. Thus, to study the saddle connections of $\Sigma$ it suffices to study its core. Although the interior $\ics$ of the core may be empty, its boundary $\dcs$ is never empty, and in particular is a polygon composed of a finite union of saddle connections. Thus, any translation surface with poles has at least some saddle connections. More precisely, any surface $S\in\Hh(a_1',\cdots,a_m',-b_1',\cdots,-b_n')$ must have at least \eq{N_{\text{min}} \equiv 2g'-2+m+n \label{eq:taharlower}} saddle connections, and moreover a surface has only this minimum number of saddle connections if and only if its core has empty interior \cite{tahar:poles}. 

\begin{figure}
\begin{center}
\includegraphics[scale=1.1]{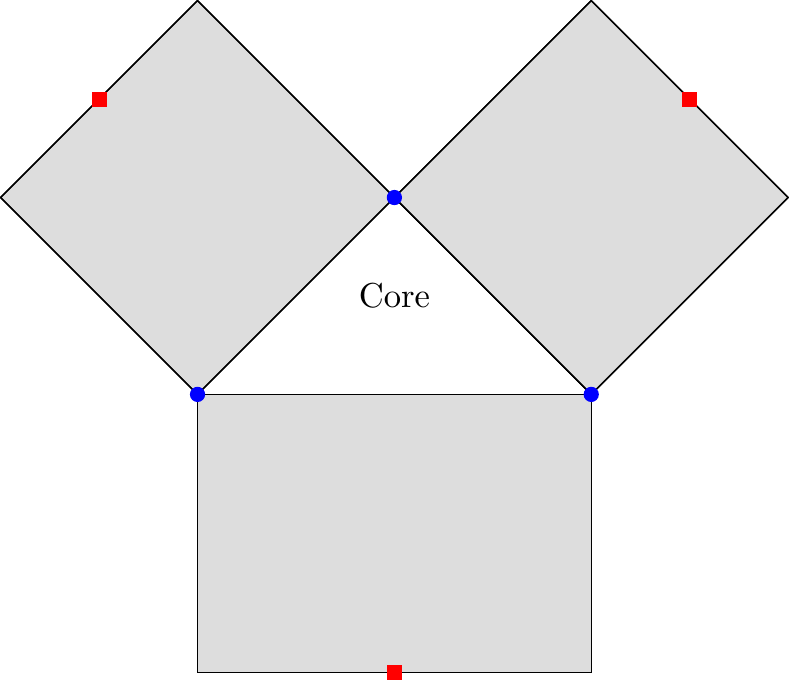}
\caption{A sketch of a flat surface with poles. Blue dots represent the zeroes of $\l$, and its poles are red squares. The center of the figure is the polygonal core where saddle connections are allowed to live, and the shaded regions are the infinite half-spaces behind the geodesic horizons.}
\label{fig:core}
\end{center}
\end{figure}

We have argued that all saddle connections live in $\cs$. On the other hand, most geodesics on $\Sigma$ will leave the core. Each pole of $\l$ sits at the end of an infinite half-space: either a half-cylinder for a simple pole or a half-plane for a pole of order two or higher. Any geodesic that leaves the core will stretch infinitely far along one of these spaces before terminating at one of the poles. The infinite size of these regions renders the total area of each translation surface with poles infinite. Thus, whereas translation surfaces without poles can be visualized as a plane polygon subject to some identifications, once poles are present the picture is radically different: $\Sigma$ decomposes into a polygonal core surrounded by infinite half-spaces. We have sketched an example in Figure \ref{fig:core}. We note briefly that this configuration is heuristically similar to geodesics propagating on a black hole geometry: particle worldlines are allowed to propagate freely until they cross the black hole horizon, at which time they are forced to end at the singularity. This is why the boundary of the core was known as the ``geodesic horizon" in \cite{warner:geodesics}. 

\begin{figure}[h]
\begin{center} \includegraphics[scale=.4]{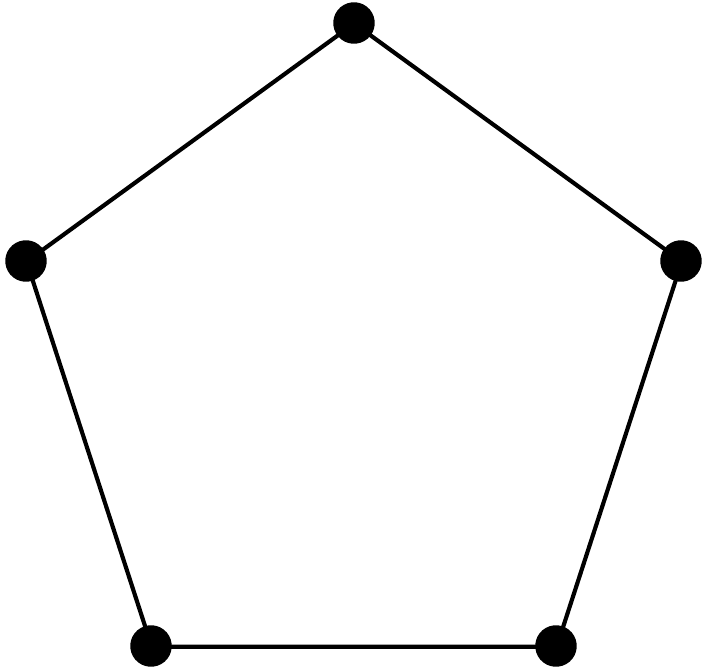}\ \includegraphics[scale = .4]{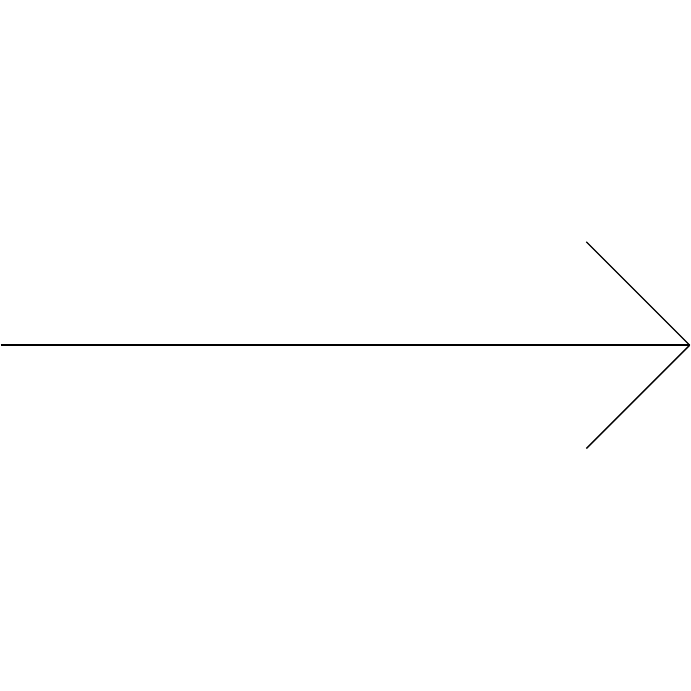}\ \ \includegraphics[scale=.4]{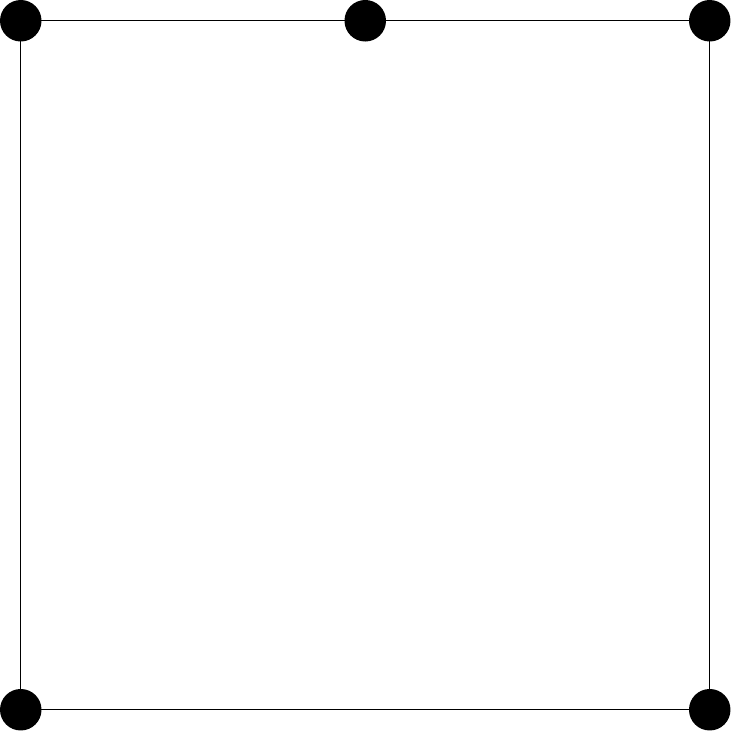}\ \ \includegraphics[scale = .4]{arrow}\ \ \includegraphics[scale=.4]{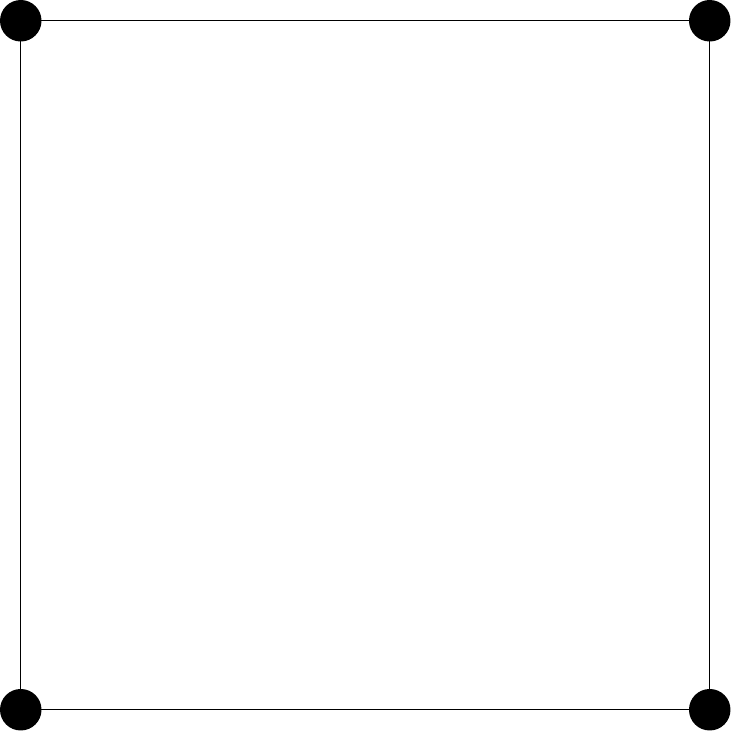} \end{center}
\caption{Change of core topology. It was proven in \cite{tahar:poles} that the core topology defines a wall-crossing structure in strata of meromorphic differentials. The first arrow shows the core undergoing a real codimension one degeneration at a wall, and the second arrow shows a further real codimension one degeneration: two zeroes collapse onto each other, leaving us in a stratum with fewer total zeroes.}
\label{fig:coreTopChange}
\end{figure}

The core is always a polygon, but the number of its sides can change as we move around in $\chb$. Imagine two consecutive saddle connections in $\dcs$, i.e. two saddle connections meeting at the same zero. These two saddles will meet at the middle zero with some relative angle. At real codimension one in the ambient stratum, we can tune the angle until the two saddles are colinear; this is sketched in the left two panels of Figure \ref{fig:coreTopChange}. If we continue deforming past this real codimension one condition, we will end up with a core with different polygonal topology. It was proven in \cite{tahar:poles} that this core topology change defines a natural wall-crossing structure on $\chb$, where the walls\footnote{In \cite{tahar:poles}, the walls were called the discriminant locus.} $\W\subset\chb$ consist of all surfaces in $\chb$ with two colinear saddles in $\dcs$, and the chambers $\C_i$, defined to be the connected components of $\chb\setminus\W$, correspond to sets of surfaces with the same core topology.

In \cite{tahar:poles}, a distinction is drawn between chambers of bounded type and those of unbounded type. In chambers of bounded type, there are only a finite number of saddle connections. Moreover, in addition to the lower bound in \eqref{eq:taharlower}, an upper bound on the number of saddle connections on a surface in a chamber of bounded type is provided; the actual value of the upper bound is given in Theorem 2.5 of \cite{tahar:poles}, but will not be relevant for our purposes. On the other hand, chambers of unbounded type consist of surfaces with infinitely many saddle connections. In surfaces belonging to chambers of unbounded type, the interior of the core decomposes into a finite number of finite volume components, each of which consists of an infinite number of saddle connections. The simplest example of such a component is a finite volume cylinder, known as an ``invariant component."   

To count saddle connections on translation surfaces with poles, especially those surfaces in chambers of unbounded type, we will need to use a set theoretic quantity known as the Cantor-Bendixon (CB) rank described, e.g., in \cite{kechris:book,aulicino:cbrank}. Consider a set $S\subset\RR$, and let $S^*$ be the set obtained from $S$ by removing all of its isolated points. We then define $S^{*(2)}$ to be the set obtained by removing the isolated points of $S^*$, and so forth. We define the CB rank $k(S)$ to be the smallest $k\in\NN$ such that \eq{S^{*(k)} = S^{*(k+1)} \ .\label{eq:cbDef}} Before applying CB ranks to saddle connections, we will give a few simple examples. First, consider a finite set $S$. All of its points are isolated, and $S^*$ is the empty set. There are no more points to remove, so $S^{*(2)}$ is also the empty set, and we have $k(S)=1$. Now take $S$ to be a convergent sequence $\{x_i\}$ converging to a point $x\in\RR$. The only nonisolated point is $x$, so $S^*$ is the singleton $\{x\}$. Then $S^{*(2)}$ is the empty set, and $k(S)=2$. This same argument goes through unmodified if $S$ is a finite collection of convergent sequences. On the other hand, if $S$ has an accumulation point of accumulation points (and nothing else), then $k(S)=3$, and so on.

Fix a surface $\Sigma\in\chb$ in some chamber $\C$, i.e., away from the walls of the stratum, and construct the set $\Theta(\Sigma)$ of saddle connection directions on $\Sigma$. It is suggested in \cite{aulicino:cbrank} that it is natural to consider the CB rank $k(\ths)$ of $\ths$, which for simplicity we will usually write simply as $\ks$. Although it has not been proven for surfaces of rank greater than two, we conjecture that each surface in a fixed chamber will have the same CB rank, so we will frequently refer to the CB rank $k_\C$ of some chamber $\C$. This conforms to our physical expectations, and is mathematically plausible: the calculations of $\ks$ in \cite{aulicino:cbrank} are based entirely on $\dcore(\Sigma)$, which is constant in each chamber, so it is sensible to suspect that the rank should be constant in each chamber as well.

\begin{figure}
\begin{center}
\includegraphics[scale=1.5]{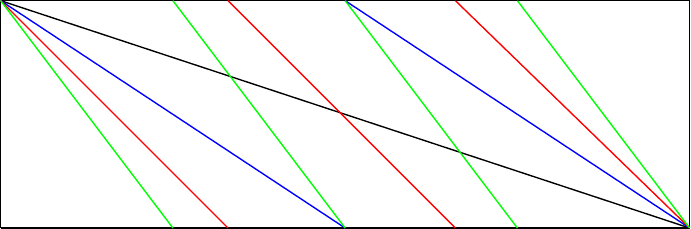}
\caption{Winding modes along a cylinder (the top and bottom of this rectangle are identified), with winding number one (black), two (blue), three (red), and four (green). As we consider geodesics with increasing winding number, the angle of the geodesics approaches the angle of the circular caps of the cylinder, giving the linear growth in \eqref{eq:k=2count}.}
\label{fig:cylinder}
\end{center}
\end{figure}

A surface with $\ks=1$ has only finitely many saddle connections, so, given that it is not on a wall, it lives in a chamber of bounded type. We thus have \eq{\ks=1 \Rightarrow N(L,\Sigma) \sim L^0 \ .\label{eq:k=1count}} Similarly, the set $\Theta(\Sigma)$ corresponding to a surface $\Sigma$ in a chamber with $k_\C=2$ should have a finite number of accumulation points. What do these accumulation points correspond to geometrically? Consider a minimal component of $\cs$, i.e., a finite volume cylinder in the core; for surfaces with $k=2$, these are the only components of the core. Each of these cylinders has an infinite number of saddle connections on it, corresponding to winding modes, as shown in Figure \ref{fig:cylinder}. The accumulation points are simply the directions of the circles at the ends of the cylinder with respect to the rest of the surface. For such surfaces, these winding modes are the only allowed infinite family of saddles, so \eq{k_\C = 2 \Rightarrow N(L,\Sigma) \sim L \ .\label{eq:k=2count}}

For chambers with $k_\C>2$, there is not a similar intuitive picture for the growth of saddle connections. However, an unpublished conjecture of Aulicino, Pan, and Su describes the growth rate for these chambers \cite{aulicino:conjecture}. They conjecture that \eq{k_\C \ge3 \Rightarrow N(L,\Sigma) \sim L\left(\log{L}\right)^{k_\C-2} \ .} Adding in \eqref{eq:k=1count} and \eqref{eq:k=2count}, we have \eq{N(L,S) \sim \left\{ \begin{array}{cc} L^0 \ ,& k_\C=1\\ L\left(\log{L}\right)^{k_\C-2}, & \text{otherwise}\end{array}\right. \label{eq:davidConjecture}.} This conjecture should be understood as holding pointwise, in the sense of \eqref{eq:masur}. For instance, for every $\Sigma$ in a chamber with $k_\C=2$ the conjecture holds that there exist positive constants $c_1,c_2$ such that \eq{c_1L \le N(L,\Sigma) \le c_2L \ .} \eqref{eq:davidConjecture} is easy to verify for $k_\C=1$, and has been proven rigorously for $k_\C=2$, essentially by formalizing the winding mode argument above, but remains conjectural for $k_\C\ge3$ \cite{aulicino:conjecture}. In this paper we give no examples with $k_\C\ge3$, but such chambers will exist in generic theories. 

As mentioned above, translation surfaces with poles have infinite area. This prevents us from constructing substrata $\chob$ for meromorphic $\beta'$, as was done for holomorphic strata above. However, an alternate construction, known as a projectivized stratum, was introduced in \cite{tahar:twoStar}. The projectivized stratum $\pchb$ is obtained from $\chb$ by identifying surfaces in $\chb$ related by complex scaling. We therefore have \eq{\dim_\CC \pchb = \dim_\CC \chb - 1 \ .} Importantly, whereas in the construction of $\chob$ we had a canonical way of selecting representatives from each equivalence class of surfaces everywhere in moduli space, once we are considering surfaces with poles the choice of representative must be decided for each chamber separately. Several examples are worked out in \cite{tahar:twoStar}; in general, we select representatives by normalizing the length of one of the saddle connections in $\dcs$. 

\section{BPS Geodesics for $A_1$ Class S Theories}\label{sec:geoA1}
So far, we have argued that BPS hypers in $A_1$ class S theories are equivalent to saddle connections on either the UV curve $C$ or, equivalently, the Seiberg-Witten curve $\Sigma$, and described some results in the math literature relevant to the counting of saddle connections. In this section, we will apply these techniques to obtain quantitative estimates for the asymptotic BPS spectrum in a variety of theories. First in \S \ref{sec:stratifying} we will describe the structure of the Coulomb branch in terms of strata, and in so doing find the stratum associated to a point in the Coulomb branch. We will then apply this framework to several examples, finding explicit results for the asymptotic degeneracy of BPS states. In \S \ref{sec:holoExamples} we will discuss non-Argyres-Douglas SCFTs, whose quadratic differentials are holomorphic, for which we can use \eqref{eq:masurQuad} and \eqref{eq:SVquadSaddle} to find quadratic asymptotics for the BPS spectrum. Then in \S \ref{sec:meroExamples} we will discuss the more general, but correspondingly more difficult, case of theories with meromorphic $\l$, for which \eqref{eq:davidConjecture} describes the spectrum of saddle connections on the SW curve; for these theories, we will provide a recipe for dealing with a general theory, but will only work out explicitly one example, pure $\N=2$ SYM with SU(2) gauge group.\footnote{A geodesic perspective on the BPS spectrum of this theory was considered in \cite{vafa:geodesics,warner:geodesics,rabin:geo}.  However, these papers were written before most of the results discussed in \S \ref{sec:mero} were known, and so could not take advantage of the close relationship between these questions and dynamical systems.}

\subsection{Stratifying the Coulomb Branch}
\label{sec:stratifying}
As discussed in \S \ref{sec:max}, to specify an $A_1$ class S theory, we fix a punctured genus $g$ Riemann surface $C$, along with the maximal orders ${b}_j$ of the poles of a quadratic differential $\phi$ at these punctures and certain coefficients of singular terms in $\phi$. We assume that we are at generic points in moduli space, so that the poles have orders exactly $b_j$. These integers then fix the singular part of a family of strata of the form $\Q(\cdots,-{b}_j)$. \eqref{eq:RRquad} fixes the sum of the orders ${a}_i$ of the zeroes, but not the number of zeroes. For instance, it is conceivable that, inter alia, $\phi$ has either all simple zeroes, or one zero of large order, and indeed there exist loci in moduli space where both of these possibilities are realized. 

However, at generic points in moduli space, we expect $\phi$ to have only simple zeroes;\footnote{This was mentioned in \S \ref{sec:max}; the goal of this subsection is to rephrase this intuition in the language of strata and saddle connections.} indeed, in the mathematics literature, quadratic differentials with only simple zeroes are known as ``GMN differentials" \cite{aulicino:cbrank,bs:quadStab}. Therefore, at most points in moduli space $\phi$ lives in the stratum $\Q(1^M,-{b}_j)$, where \eq{M \equiv \sum_{i=1}^m {a}_i = 4g-4 + \sum_{j=1}^n {b}_j \ .}  Under the canonical double cover, this stratum maps into an abelian stratum of the form $\Hh\left(2^M,-b_j'\right),$ where the number of poles and their orders $b_j'$ are determined by the rules laid out in \S \ref{sec:rels}. These abelian differentials are defined on a Riemann surface whose genus $g'$ depends on the genus $g$ of $C$ and the order of the poles of $\phi$. 

If $\phi$ has poles of higher order, then the abelian stratum will be a stratum of meromorphic differentials, and by \cite{tahar:poles} will have walls at real codimension one. At these walls, two saddle connections in $\dcore(\Sigma)$ will become colinear and then merge into a single long saddle connection. This is quite reminiscent of two BPS states merging into one stable bound state as we cross a wall. For this picture to be accurate, we would need to associate the length $L$ of each geodesic with the mass $M$ of a BPS state,\footnote{Really, the length is half of the mass, since as explained in \S\ref{sec:max} we need to include the mirror images of our curves on the two sheets of $\Sigma$. We will neglect this factor of 2.} and the direction $\theta$ of a geodesic with the phase of the central charge $Z=Me^{i\theta}$ of the corresponding BPS state. The identification of the mass with the length was already seen in \eqref{eq:massintegral}, and similarly the relationship between the direction of a geodesic and the phase of a central charge is given in \eqref{eq:geodesicAngle}. Thus we see that the notion of wall crossing natural in the geodesic framework lines up exactly with the physical picture of wall crossing.

Generically, the canonical double cover is not a surjection: the images of most quadratic strata will be only a subset of the corresponding abelian stratum. Therefore, not all of the walls $\calW$ of $\Hh\left(2^M,-b_j'\right)$ will be BPS walls in the physical moduli space. Instead, the intersection $\calW\cap\frakC \B$ of the geodesic walls in the abelian stratum with the image of the Coulomb branch will give physical walls. As required of walls in moduli space, this locus has real codimension one in the Coulomb branch. Note, however, that these might not be the only walls in the theory. Walls that do not change the asymptotic growth rate of BPS states in the theory would be invisible to this point of view, but should exist in general. We will presently see an example of this. 

At complex codimension one in $\B$, we can have two simple zeroes of $\phi$ condense into one double zero, and therefore end up in the stratum $\Q(2,1^{M-2},-{b}_j)$. Similarly, a zero can collide with a simple pole. Physically, these events characterize the sublocus of moduli space where BPS states become massless, sometimes known as the discriminant locus. BPS states can only become massless on the walls of marginal stability, as a vanishing central charge aligns with all other central charges. So, zeroes can only condense or collide with simple poles in walls of $\B$. Thus, class S theories whose quadratic differentials have zeroes generically have wall crossing, even for theories with holomorphic differentials, for which the arguments of \cite{tahar:poles} do not apply. For holomorphic theories, \eqref{eq:masurQuad} gives the asymptotic count at all points in moduli space, so any walls that are present do not change the asymptotic spectrum.

To avoid subtleties involving bound states at threshold, we will restrict ourselves to points of $\B$ away from the walls; however, we will briefly return to the walls in the conclusion. Thus, we will be able to focus on a single stratum of the form \eq{\Q\left(1^M,-b_1,\cdots,-b_n\right).\label{eq:thestratum}} In the remainder of this section, we will apply this framework to obtain explicit estimates for the growth rates of BPS state counts for a variety of theories.

\subsection{Non-Argyres-Douglas SCFTs}
\label{sec:holoExamples}
We will first discuss theories for which the relevant differentials are holomorphic. This means in particular that $\phi$ is allowed to have simple poles, but no higher-order poles. The results of this section therefore apply to all non-Argyres-Douglas SCFTs of type $A_1$. 

We are thus led to consider an infinite family of theories: we fix the genus $g$ of $C$ and the number $n$ of simple poles of $\phi$. At generic points in moduli space, we are led to consider choices of $\phi$ in the stratum \begin{subequations} \eq{\Q\left(1^n,-1^{n+4}\right)\label{eq:Qsimplepolesg=0}} for $g=0$, or in the stratum \eq{\Q\left(1^{n+4g-4},-1^n\right)\label{eq:Qsimplepolesg>0}} \label{eq:Qsimplepoles}\end{subequations} otherwise.\footnote{$g=0$ is not really a special case. If we plug $g=n=0$ into \eqref{eq:Qsimplepolesg>0}, we find $\Q(1^{-4})$. We can't have a negative number of zeroes, so these are just simple poles. We have split the two up to avoid this possibly confusing notation.} The stratum $\Q(1,-1)$ is empty \cite{masur:prescribed}, so we exclude the case $g=n=1$.

We now want to study the spectrum of BPS hypers in these theories, and in particular to compute the number $N(M,u)$ of BPS hypers of mass less than $M$ at $u\in\B$. BPS hypers are given by by saddle connections on the SW curve $\Sigma$, or equivalently by saddles on the UV curve $C$, but it is easier to count these saddle connections on $C$. When $C$ has unit area we can use Siegel-Veech\footnote{We are justified in using Siegel-Veech since the moduli space of holomorphic theories is all of \eqref{eq:affine}, and thus we are considering generic points in the stratum.} for holomorphic quadratic differentials to find that \eq{\lim_{M\to\infty}\frac{N_{g,n}(M,u)}{\pi M^2} = c_{\Q(\b)} \text{ for almost all $u$ away from the walls} \ ,\label{eq:holoAnswer}} where $\Q(\b)$ is the stratum indicated in \eqref{eq:Qsimplepoles} and $c_{\Q(\b)}$ is its SV constant. When $C$ does not have unit area, we can use \eqref{eq:Nscaling} to generalize \eqref{eq:holoAnswer} to other choices of total scale. This rescaling cannot force us to cross a wall, as real multiplication cannot cause any two central charges to align. We can relax the genericity condition by applying \eqref{eq:masurQuad} to find that \eq{N_{g,n}(M,u) \sim M^2\ \forall\ u\in\B \text{ away from the walls} \ . \label{eq:holomasur}}

For most values of $g$ and $n$, $\phi$ and $\l$ will have zeroes, and therefore we can have zeroes condense. As discussed above, when this happens, we have light BPS states, and therefore must have walls. However, as long as we restrict ourselves to only be in one connected component of $\Q(\b)$, the growth rate is given almost everywhere by \eqref{eq:holoAnswer} (and everywhere by \eqref{eq:holomasur}) in any chamber. Thus, whatever wall crossing we have must not affect the asymptotics of the spectrum. The particular charges which support states may change, but not the asymptotic count. Mathematically, the necessity of `almost everywhere' in \eqref{eq:holoAnswer} is demonstrated by the existence of `Veech surfaces' (see, e.g., \cite{lelievre:veech}).

The examples where neither $\phi$ nor $\l$ have zeroes merit special attention. There are two such examples. The first occurs when $C$ is a torus with no punctures at all, i.e. $\N=4$ $SU(2)$ SYM. For $\N=4$ SYM, there are no zeroes or marked points to pick out, and therefore it is no longer natural to count saddle connections; physically, this corresponds to the absence of hypers. Instead, as discussed in \S\ref{sec:max} we should count families of parallel closed geodesics on the torus \cite{warner:geodesics,brand:geodesics}. Thus, inserting \eqref{eq:SVtorus}, we have \eq{\lim_{N\to\infty}\frac{N_{\N=4}(M,u)}{\pi M^2} &= \frac{6}{\pi^2} \text{ for almost all $u\in\B$}\ \label{eq:n=4answer}.}
The physical spectrum consists of dyons of charges $(p,q)$ for all coprime $p$ and $q$ \cite{sen:S,porrati:S}. Counting such states with mass at most $M$ gives exactly the result of \eqref{eq:n=4answer}, so the geodesic point of view reproduces the known results \cite{warner:geodesics,zorich:flat,mz:slagCounts}.

The second example with no zeroes has $C$ a sphere with four simple punctures, corresponding to $\N=2$ $SU(2)$ SYM with $N_f=4$. In this case, we have $\phi\in\Q(-1^4)$, and therefore $\l\in\Hh(0^4)$. \eqref{eq:SVquadSaddle} then gives us \eq{\lim_{N\to\infty}\frac{N_{N_f=4}(M,u)}{\pi M^2} &= c_{\Q(-1^4)} \text{ for almost all $u\in\B$}\ \label{eq:nf=4answer}.} The BPS spectrum for this theory is also known \cite{sw,ferrari:finite}, and can again be seen to have quadratic asymptotics.

Finally, we turn to vector multiplets. There is nothing new to say here: as explained in \S\ref{sec:holo}, the situation of closed geodesics on $C$ that lift to closed geodesics on $\Sigma$ closely parallels that of saddle connections on $C$. (In particular, there is a SV constant that controls growth almost everywhere, and \eqref{eq:masur} gives us the asymptotic growth rate everywhere.)

\subsection{Meromorphic Theories}
\label{sec:meroExamples}
We will now move on to the general case, in which $\phi$ has poles of order at least 2. In this case, \eqref{eq:masurQuad} and \eqref{eq:SVquadSaddle} do not apply, and instead we must use \eqref{eq:davidConjecture}, which necessitates computing the CB rank of each chamber in the Coulomb branch, to count saddles on $\Sigma$. The chambers in moduli space can be determined using $\l$ fairly straightforwardly: one computes the periods of $\l$ over all the cycles of $\Sigma$, and simply solves for the loci in moduli space where the central charges of two BPS states align. 

On the other hand, it is sometimes possible to find the chambers from a purely geodesic perspective, without ever computing periods of $\l$. Instead, by simply computing the allowed core topologies in some stratum $\Hh(\b')$, we can find all of the allowed chambers and moreover can find detailed criteria for what types of chambers can be adjacent to each other in the full stratum; this was done for strata of the form $\ph(a,-a)$ in \cite{tahar:twoStar}. This analysis uses a very different set of techniques than the usual procedure, being predominantly combinatoric in nature, and thus is an interesting alternate perspective.

Once we have all of the chambers of the Coulomb branch, techniques similar to those of \cite{aulicino:cbrank} should allow us to compute the CB rank of each chamber, and from there \eqref{eq:davidConjecture} gives us the asymptotic spectrum of hypers. (As mentioned in \S \ref{sec:mero}, the spectrum of closed geodesics for meromorphic differentials is not well understood, and therefore we cannot say anything about the spectrum of vectors.) However, doing this is quite difficult in practice. In \S \ref{sec:n=2} we will consider an explicit example, $\N=2$ SU(2) SYM, and study it in detail, finding both the structure of the moduli space and the asymptotic BPS spectrum in each chamber from a geodesic point of view. We will then make some statements about a much larger class of meromorphic theories in \S \ref{sec:misc}. 

\subsubsection{Pure $\N=2$ SU(2) SYM}
\label{sec:n=2}

 Let us study the BPS states of the pure SU(2) theory. First we will confirm the picture of the Coulomb branch laid out in \S \ref{sec:stratifying}.  As discussed in \S \ref{sec:max}, for this theory we take $C$ to be a genus 0 Riemann surface with two order three punctures. We thus anticipate that, away from the walls, $\phi$ should be in the stratum $\Q(1^2,-3^2)$, and therefore that $\l$, defined on $\Sigma\subset T^*C$, should be in the stratum $\Hh(2^2,-2^2)$. We have that \eq{\phi = \l^2  = \left(\frac{\Lambda^2}{z^3}+\frac{u}{z^2}+\frac{\Lambda^2}{z}\right)\, dz^2.} This differential has simple zeroes at \eq{z_\pm = \frac{-u\pm\sqrt{u^2-4\Lambda^4}}{2\Lambda^2}} and triple poles at $z=0,\infty$, so $\phi\in\Q(1^2,-3^2)$, as expected. On the other hand, if we had instead studied the singularity structure of $\l$, we would have found double zeroes at $z_\pm$, and double poles at $z=0,\infty$, again as expected. Finding the walls explicitly is in general hard, but it is quite easy to find the discriminant locus in this case. It occurs when $z_+=z_-$, i.e. at $u=\pm 2\Lambda^2$ -- the monopole and dyon points! Analyzing $\l$ at these points, we find one simple zero with multiplicity two, leaving us in the stratum $\Hh(1^2,-2^2)$. 

\begin{figure}
\begin{center}
\includegraphics[scale=1]{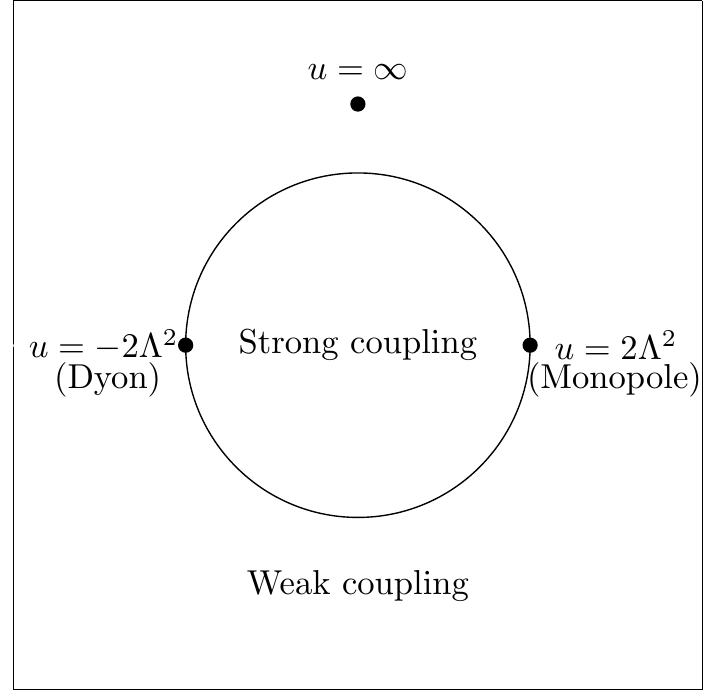}
\caption{The Coulomb branch of pure $\N=2$ SYM with SU(2) gauge group.}
\label{fig:uPlane}
\end{center}
\end{figure}

We would now like to actually count BPS states in this theory. Of course, the spectrum is known exactly, as is the structure of the Coulomb branch, which we have reproduced in Figure \ref{fig:uPlane}. In the strong coupling region we have a dyon of charge $\pm(2,1)$ and a monopole of charge $\pm(0,1)$, and in the weak coupling region we have a W-boson of charge $\pm(2,0)$ and dyons of charge $\pm(2n,1)$ with $n\in \ZZ$. Therefore, the asymptotic count of BPS hypers of mass at most $M$ is given by \eq{N(M,u) \sim \left\{\begin{array}{cc}M^0,&u\in\text{ strong coupling region}\\M^1,&u\in\text{ weak coupling region}\end{array}\right..} We therefore expect the intersection $\B\cap\Hh(2^2,-2^2)$ to have two chambers, one with $k=1$ and the other with $k=2$.

To reproduce these results from a geodesic point of view, we will play a series of tricks. First, there is an involution of $\Sigma$ under which $\lambda$ is invariant, namely $(x,z)\mapsto (-xz^2,1/z)$; another way of thinking about this is swapping the north and south poles of $C$ and then modifying $x$ so that $\lambda=x\,dz$ is invariant. As discussed in \S\ref{sec:lowNRG}, we now quotient $\Sigma$ by this symmetry, as was done in \cite{warner:geodesics}, to end up in the stratum $\Hh(2,-2)$. By \eqref{eq:meroabeldim}, this stratum has complex dimension two, whereas the Coulomb branch has dimension one. Therefore, to locate the Coulomb branch, we must remove one complex degree of freedom from the stratum. We do this by projectivizing, leaving us with the stratum $\ph(2,-2)$. Physically, this identification amounts to fixing the complexified scale $\Lambda$. It might seem surprising that we only need to fix the coefficient of the double pole in $\lambda$, but not the residue at this point; however, the latter automatically vanishes in $\Hh(2,-2)$, by the residue theorem. This is already sufficient to find the Coulomb branch; for more general theories, fixing parameters and dynamical scales would force one to carve out a higher codimension slice of a stratum.

\begin{figure}
\begin{center}
\includegraphics[scale=1]{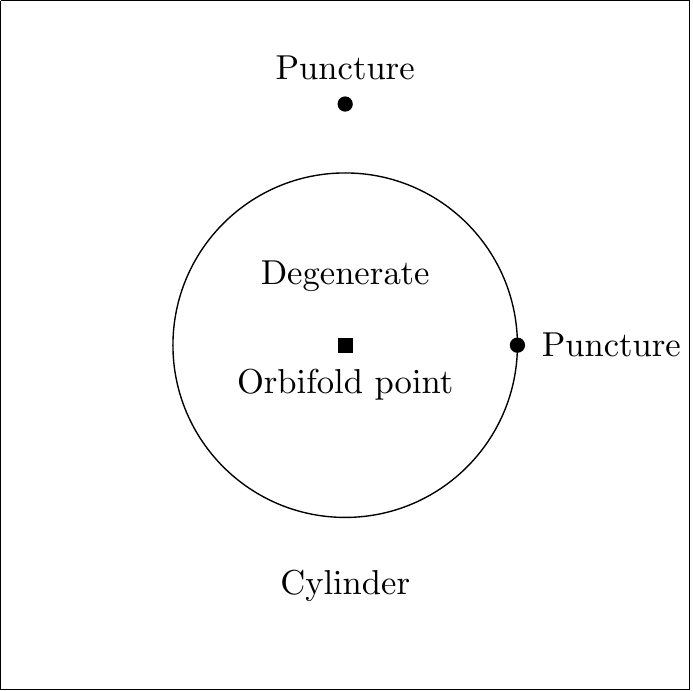}
\caption{The structure of $\ph(2,-2)$, as derived in \cite{tahar:twoStar}.}
\label{fig:phtt}
\end{center}
\end{figure}

The wall-crossing structure of $\ph(2,-2)$ was worked out in Proposition 4.9 of \cite{tahar:twoStar} using combinatorial information about flat surfaces, rather than detailed periods of the SW differential; the results are summarized in Figure \ref{fig:phtt}, which is clearly similar, but not identical, to the picture of the Coulomb branch in Figure \ref{fig:uPlane}. However, we will see that it is exactly the correct geometry to reproduce the spectrum of $\N=2$ SYM. We'll start with the growth rates in each chamber, and then work out how they fit together to give the space shown in Figure \ref{fig:uPlane}. We begin with the degenerate chamber. In degenerate chambers, as defined in \cite{tahar:twoStar}, the core has no interior, and we have only the saddle connections that comprise the boundary of the core itself; we therefore have $k=1$ for this chamber. The other chamber is of cylinder type. Not surprisingly, in a chamber of cylinder type, the core is a cylinder, and we have $k=2$. \eqref{eq:davidConjecture} then gives us the growth rates of saddle connections in the two chambers: \eq{N(L,S) \sim \left\{\begin{array}{cc}L^0,&S\in\text{ chamber of degenerate type}\\L^1,&S\in\text{ chamber of cylinder type}\end{array}\right..} We therefore identify the chamber of degenerate type with the strong coupling region and the chamber of cylinder type with the weak coupling region.

\begin{figure}
\begin{center}
\includegraphics[scale=1]{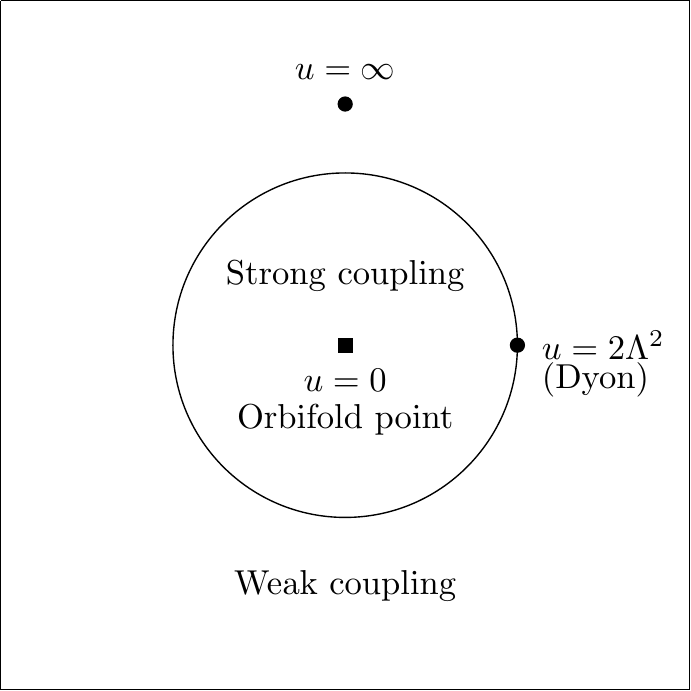}
\caption{The quotient of the Coulomb branch of $\N=2$ SYM by $u\mapsto-u$.}
\label{fig:uPlaneQuotient}
\end{center}
\end{figure}

Thus, we see that the geodesic picture reproduces the BPS spectrum away from singular points in moduli space. The singular points in Figure \ref{fig:phtt}, however, appear not to match with those in Figure \ref{fig:uPlane}. To understand the relationship between the two, it is very instructive to start with the orbifold point; everything else will follow. It is well-known that the theory has a spontaneously broken $\ZZ_2$ R-symmetry under which $u\mapsto-u$. Under this symmetry, as well as the diffeomorphism $z\mapsto -z$, we have \eq{\phi \mapsto \parens{-\frac{\Lambda^2}{z^3} - \frac{u}{z^2} - \frac{\Lambda^2}{z}}dz^2 = -\phi \Rightarrow \lambda \mapsto \pm i\l \ .}We see that the action of this R-symmetry simply rotates $\l$. But we are working in $\ph(2,-2)$, so these points are identified!  Thus $\ph(2,-2)$ is not really the $u$-plane, but is instead the quotient of the $u$-plane by this global symmetry; we have drawn this quotient in Figure \ref{fig:uPlaneQuotient}. 

Now everything falls into place. The strong coupling region contains the point $u=0$, which is an order 2 fixed point under $\ZZ_2$; this is the promised orbifold. Next, we move on to the monopole and dyon points. Previously, we had two singularities at $u=\pm2\Lambda^2$, through which ran the wall of marginal stability. Now, however, these two points are identified, and the wall only runs through one singularity, again as advertised. Finally, the point at infinity never existed in the physical theory in the first place, so we need the puncture at infinity in order to reproduce the $\ZZ_2$ quotient of $\B$. Thus, the geodesic picture has completely reproduced the known physics, up to a $\ZZ_2$ quotient!

\subsubsection{General Results for Meromorphic Theories}
\label{sec:misc}
Although we have no further concrete examples of theories with meromorphic differentials whose spectra we have studied, there are still general statements we can make, purely based on the strength of existing mathematical results. For example, \cite{tahar:poles} provides a complete classification of strata for which no chamber has infinitely many saddle connections. In particular, half-translation surfaces in all such strata have genus zero, so any $A_1$ class S theory whose UV curve is of genus $g>0$ lives inside a stratum that has at least one chamber with infinitely many saddle connections. (However, this chamber might not overlap with the Coulomb branch.)

Moreover, Corollary 2.4 of \cite{tahar:poles} gives a sharp criterion for whether or not a fixed stratum of genus zero meromorphic quadratic differentials admits a chamber with infinitely many saddle connections. The stratum $\Q(a_1,\cdots,a_m,-b_1,\cdots,-b_m)$ admits such a chamber unless its singularity structure $\{a_1,\cdots,a_m,-b_1,\cdots,-b_m\}$ cannot be decomposed into two sets, each of which contains at least one zero, such that the sum of the orders of the zeroes and poles of each set is $-2$. For instance, the stratum $\Q(1^2,-3^2)$ corresponding to pure SU(2) SYM can be decomposed as $\{\{1,-3\},\{1,-3\}\}$, which guarantees that there exist surfaces in the stratum with infinitely many saddle connections. On the other hand, the stratum $\Q(1,-1,-4)$ admits no such decomposition, and so the theory corresponding to this stratum (two free hypers, also known as the Argyres-Douglas $X_4$ theory) never has infinitely many BPS hypers; this observation generalizes to any AD $X_N$ or $Y_N$ theory.  

In addition, there are some statements that we can make for individual theories without doing any extra work. For instance, it is explained in \cite{donagi:integrable,warner:geodesics} that $\N=2^*$ $SU(2)$ SYM can be studied in the stratum $\Hh(2,-1^2)$, after taking a quotient of the Seiberg-Witten curve. Proposition 8.3 of \cite{tahar:poles} states that each chamber in this stratum is of unbounded type, so away from the walls each point in the moduli space of this theory has infinitely many BPS states; this confirms the suspicions of \cite{longhi:invt}. The same argument applies to any theory for which $C$ has exactly one second order pole and any number of simple poles, i.e. the doubly infinite tower of theories $\Q(1^{4g+N-2},-1^N,-2)$; the only exception is the stratum $\Q(-1^2,-2)$, which via the above criterion has no surfaces with infinitely many saddles.

\section{Conclusion}\label{sec:conclude}

In this paper, we have employed ideas from the study of flat surfaces to investigate the asymptotic growth of BPS state counts in $A_1$ class S field theories. We conclude by suggesting several avenues for continued research in this direction.

One natural course is to generalize these results away from theories of $A_1$ type, i.e. to extend the framework laid out here to the rest of class S. Then instead of simply counting geodesics on Riemann surfaces, we would have to contend with string webs, i.e. webs of geodesics glued together in particular ways. An analog of \eqref{eq:davidConjecture} for string webs would give us asymptotic results for the BPS spectrum in each chamber of all class S theories.

Even for $A_1$ theories, there are many interesting examples that this formalism should help with. A good example is $\N=2^*$ $SU(2)$ SYM, which as mentioned above should be studied in the stratum $\Hh(2,-1^2)$. Although the BPS geodesic prescription was used in \cite{warner:geodesics} to make significant progress towards understanding the spectrum of this theory, and the spectrum is in principle determined at all points in moduli space by the spectrum generator computed in \cite{longhi:thesis,longhi:invt}, it is nevertheless the case that our understanding of even the qualitative behavior of these BPS state counts throughout moduli space remains incomplete. However, given the recent advances in the study of flat surfaces with poles, it seems likely that the asymptotics of the spectrum can be determined readily. In particular, projectivizing the stratum and fixing the complex structure modulus of the elliptic curve $C$, i.e. working at a fixed point in parameter space,\footnote{Projectivizing, which should always be thought of as fixing a dimensionful parameter, allows us to pin down the mass parameter, and fixing $\tau$ then removes the remaining complex dimension.} leaves us with a one-dimensional Coulomb branch, at which point combinatorial arguments along the lines of those used in \cite{tahar:twoStar,aulicino:cbrank} should suffice to determine all of the allowed chambers and their adjacencies,\footnote{Of course, one could also find the walls and chambers by analyzing the periods of $\l$.} as well as the CB rank of each chamber; \eqref{eq:davidConjecture} would then allow us to determine the asymptotic BPS spectrum in each chamber.

Certainly, one hopes for a physical explanation of a number of the facts that have been discussed in this paper. Perhaps chief among these is the universal quadratic asymptotics in non-Argyres-Douglas $A_1$ class S SCFTs. What is physically special about them, and do other theories share this property? (One possible answer to this question is the notion of `completeness' \cite{vafa:complete}; indeed, we note that this could allow one to physically reproduce the mathematical proof via averaging.) Another observation demanding a physical explanation is the large genus asymptotics discussed in \cite{eskin:largeG,zagier:largeG0,sauvaget:volumes,aggarwal:largeG2,aggarwal:largeG,zagier:largeG}, which implies asymptotics for $SU(2)^N$ theories in the large $N$ limit, and the recursive results of \cite{zagier:largeG}.

Additionally, it would be interesting to use geodesic counts to study BPS state counts on walls of marginal stability. Physically, this involves subtleties with bound states at threshold, and correspondingly the counts need no longer be integers. Indeed, Joyce and Song \cite{joyceSong} have studied this problem from a mathematical perspective and defined fractional invariants on walls.

It could be interesting to adapt ideas surveyed in this paper to the study of black hole microstate counting, and vice versa. In the former direction, we first note the obvious holographic implications of this work. Secondly, the idea of averaging over points in parameter or moduli space in order to discern generic asymptotic growth rates of BPS state counts applies more generally than to $A_1$ class S field theories. For instance, this philosophy was applied in \cite{filip:counting} to a problem that was subsequently connected to BPS state counts of M-theory on K3 \cite{mz:slagCounts}. In the latter direction, we note that in certain gravitational contexts it is known that asymptotic growth rates of BPS states are controlled by the Bekenstein-Hawking entropy at the attractor point in moduli space and are invariant under wall crossing \cite{sen2007walls}; this is reminiscent of the behavior of BPS state counts in non-Argyres-Douglas theories in this paper, and it would be interesting if lessons from the gravitational problems (e.g., the utility of automorphic forms) could be applicable to field theories.

We conclude by offering a short wish list for future research into translation surfaces with poles. In \S\ref{sec:mero} we used two important conjectures: the count of saddle connections in \eqref{eq:davidConjecture}, and the constancy of the CB rank in each chamber. Attempts to prove \eqref{eq:davidConjecture} are currently underway \cite{aulicino:conjecture}, but to the best of our knowledge this is the first mention of the possibility that the CB rank is constant in each chamber. It would be interesting if this physically-motivated conjecture could be verified rigorously.

For theories with holomorphic differential, \eqref{eq:masur} establishes quadratic asymptotics for both hypers and vectors. Conversely, while \eqref{eq:davidConjecture} gives the corresponding prediction for the spectrum of hypers for meromorphic theories, to the best of our knowledge there is no analogous result for closed geodesics, and therefore we can make no statements about the spectrum of BPS vector multiplets in these theories. Thus, it would be excellent if the conjecture in \eqref{eq:masur} were generalized to also apply to closed geodesics. The obvious guess, $c_1L\log{L}^{k-2}\le N'_{\text{closed}}(L,\Sigma) \le N'(L,\Sigma) \le c_2L\log{L}^{k-2},$ cannot be correct, since pure SU(2) SYM has only finitely many BPS vectors in the weak coupling region.

Finally, it is tempting to conjecture that, as with $\chob$ in the case of holomorphic differentials, the study of projectivized strata would allow one to prove an SV theorem for meromorphic differentials. Given \eqref{eq:davidConjecture}, such a statement could take the form \eq{N(L,\Sigma) = c_\C L\left(\log{L}\right)^{k_\C-2}\text{ for almost all $\Sigma$ in a fixed chamber $\C\subset\pchb$}\ .\label{eq:SVconjecture}}
There is a natural $\operatorname{SL}(2,\RR)/\operatorname{SO}(2,\RR)$ action on projectivized strata, so by analogy with the holomorphic case it seems that an important step in this direction could be to search for an invariant finite-volume measure on projectivized meromorphic strata, or subsets thereof which are singled out by physical considerations.

\section*{Acknowledgements}

We thank Jayadev Athreya, David Aulicino, Laura Fredrickson, Rafe Mazzeo, Greg Moore, Andy Neitzke, Yuji Tachikawa, and Anton Zorich for interesting and helpful discussions. We also thank Jayadev Athreya for helpful comments on an earlier draft of this paper. 
S.K. acknowledges funding from a Simons Investigator Award and the National Science Foundation under grant NSF-PHY-1720397.
R.N. is funded by NSF Fellowship DGE-1656518 and an EDGE grant from Stanford University.
A.T. is supported by NSF-MSPRF grant number 1705008.

\newpage
\bibliography{Refs}

\end{document}